\documentstyle[kemulateapj]{article}

\begin{document}

\title{Multiple Scattering in Clumpy Media.\ II.\ Galactic Environments}

\author{Adolf N.\ Witt}
\affil{Ritter Astrophysical Research Center, The University of Toledo \\
   Toledo, OH 43606; awitt@dusty.astro.utoledo.edu }
\and
\author{Karl D.\ Gordon\altaffilmark{1}}
\affil{Department of Physics \& Astronomy, Louisiana State University \\
   Baton Rouge, LA 70803; gordon@dirty.phys.lsu.edu}
\altaffiltext{1}{Present address: Steward Observatory, University of
   Arizona, Tucson, AZ 85721}

\lefthead{Witt \& Gordon}
\righthead{Clumpy Dust.\ II.}

\begin{abstract}
We present and discuss the results of new multiple-scattering
radiative transfer calculations for three representative types of
galactic environments, filled with either homogeneous or two-phase
clumpy dust distributions. Extinction and scattering properties for
two types of interstellar dust, similar to those found in the average
diffuse medium of the Milky Way Galaxy (MW) and the Bar of the Small
Magellanic Cloud (SMC), are considered. The wavelength coverage
extends from 1000 \AA\ to 30,000 \AA, with particular emphasis on the
rest-frame UV. This makes these models especially applicable to
starburst galaxies and Lyman-break galaxy samples.  The examination of
the models concentrates on the study of UV/visual/near-IR reddening
effects, the wavelength dependence of attenuation, and on the changes
that arise from the transition from homogeneous to clumpy dust
distributions in different star/dust geometries. Embedded dust,
especially when clumpy, leads to saturation at fairly low reddening
values with correspondingly gray attenuation functions.  This makes
the assessment of the attenuation of the far-UV flux from starburst
galaxies difficult, if only UV/visual/near-IR data are available.
Existing data for UV-selected starburst galaxies indicate a range of
UV attenuation factors of 0--150. Our models reproduce the ``Calzetti
Attenuation Law'', provided one adopts SMC-type dust and a clumpy
shell-type dust distribution surrounding the starbursts. The average
far-UV attenuation factor for the Calzetti sample is found to be
7.4. The only relatively reliable measure for the UV attenuation
factor for single galaxies was found in the ratio of the integrated
far-IR flux to the far-UV flux, measured near 1600 \AA, requiring the
measurement of the entire spectral energy distribution of galaxies.
\end{abstract}


\section{Introduction}

The presence of dust in galaxies has a profound impact on observable
characteristics such as their apparent UV/optical/near-IR luminosity,
their spectral energy distribution, their recognizable level of
star-formation activity, and their morphology. In addition, dust
directly affects and often controls the physical conditions within the
interstellar medium of galaxies and plays a major role in interstellar
chemistry by providing the sites for the formation of interstellar
molecular hydrogen and by regulating the interstellar radiation
field. Quantitative estimates of the dust mass responsible for the
observed effects are needed in order to yield gas-to-dust ratios in
galaxies of different types, essential if we are to understand the
chemical evolution of galaxies over cosmological time scales.

Radiative transfer simulations often provide the most direct way to
arrive at a quantitative assessment of the ways that lead to observed
dust effects, provided such simulations are sufficiently realistic to
apply to actual physical systems. All such simulations involve
approximations. Historically, these have ranged from the adoption of
absorbing screen geometries (\cite{hol58}, 1975) in analogy to the
study of extinction of starlight in our Galaxy, to the consideration
of various geometries with embedded dust, including the effects of
multiple scattering (Witt, Thronson, \& Capuano 1992, hereafter
\cite{wit92b}; \cite{bru88}). According to \cite{wit92b}, a galaxy may be
looked upon as consisting of environments which are characterized by
certain geometrical relations between the attenuating dust and the
illuminating stars, For example, starburst regions may be approximated
by regions occupied by luminous stars, surrounded on the outside by
the remnants of molecular clouds from which the stars
originated. Quiescent parts of dusty disk galaxies may resemble
systems in which stars and dust are fairly uniformly mixed, or, if one
considers the same case at longer wavelengths where the dominant
sources might be K-giants, a case where the scale height of sources is
much larger than the scale height of the dust which is concentrated
toward the center or central plane. As shown by \cite{wit92b}, equal
amounts of dust in these different configurations produce very
different reddening and attenuation effects. For the analysis of
effects resulting from changes in the inclination of disk galaxies,
disk models with embedded dust have been developed, either in
plane-parallel or doubly exponential geometry (e.g.\ \cite{bru88};
\cite{byu94}; \cite{dib96}; \cite{cor96}). All
of the models mentioned so far consistently illustrate the fact that a
certain quantity of embedded dust produces much smaller attenuation
effects than a foreground screen made from the same amount of dust,
and that the inclusion of multiple scattering reduces these effects
even further by returning the scattered radiation to the optical
radiation field, albeit with an altered spectral and directional
distribution.

An approximation in common among all the models mentioned above is the
homogeneous nature of their assumed dust distributions, which are
either constant or of a monotonically varying continuous variety. In
the first paper in the present series (Witt \& Gordon 1996, hereafter
\cite{wit96}), we conducted an extensive exploration of the effects of
clumpiness on the transport of radiation through a spherical system
containing scattering dust and a single central source. This study,
together with the results of earlier work on transfer through clumpy
media (e.g.\ \cite{nat84}; \cite{boi90}; \cite{hob93}) clearly
revealed that the clumpy multi-phase structure of the interstellar
medium (ISM) leads to a significant reduction in the opacity produced
by a given distribution and mass of interstellar dust. In addition,
\cite{wit96} showed that the relationship between the effective
optical depth of a clumpy distribution and that of an equivalent
homogeneous distribution is non-linear; this leads to an effective
attenuation as a function of wavelength that differs from the opacity
law intrinsic to the dust in the system. These conclusions have been
further amplified in a recent investigation of the radiative transfer
in the clumpy environment of young stellar objects by Wolf, Fischer,
\& Pfau (1998) and general investigation of clumpy radiative transfer
by V\'{a}rosi \& Dwek (1999). Since clumpiness is a reality of the ISM
wherever it can be observed with sufficient spatial resolution, we are
introducing it into the transfer of radiation in such galactic
environments as investigated in \cite{wit92b} now.  A preliminary
version of some aspects of this work appeared in Gordon, Calzetti, \&
Witt (1997).

In addition to adding models with a clumpy two-phase structure, we
have improved several other aspects of the original \cite{wit92b}
study. Our knowledge of dust scattering properties in the UV has
greatly improved since 1992, and with some confidence we are now
modeling the transfer for 15 (out of a total of 25) bandpasses in the
wavelength range from 1000~\AA\ to 3000~\AA, compared to two in
\cite{wit92b}.  This improved UV coverage makes the current models
particularly well suited for the analysis of starburst galaxies, both
nearby and at high redshift, since these systems emit much of their
stellar energy at UV wavelengths.  Furthermore, most previous
multi-wavelength radiative transfer studies have assumed opacity
functions and scattering properties representative of average Milky
Way (MW) dust.  Even within the MW dust characteristics vary from one
environment to another (e.g.\ \cite{fit99}) and already in the local
group of galaxies there exists a considerable variety of opacity laws
(\cite{bia96};
\cite{gor98}; \cite{mis99}). In starburst
galaxies, in particular, it has been shown (\cite{gor97}) that the
opacity function intrinsic to the dust present there is much more
similar to dust found typically in the Bar of the Small Magellanic
Cloud (SMC).  We therefore model all our galactic environments with
both the average MW and SMC bar dust.  Other radiative transfer
studies which have included both MW and SMC bar type dust are those
detailed by Gordon et al.\ (1997), Takagi, Arimoto, \& Vansevicius
(1999), and Ferrara et al.\ (1999).  The models presented in the
present paper are particularly well suited to analyze the reddening
effects and the corresponding UV attenuation in Lyman-break galaxies
and galaxies seen in the Hubble deep fields (\cite{dic97};
\cite{meu97}; \cite{meu99};
\cite{saw97}; \cite{saw98}, \cite{ste99}). They 
are equally well usable for studying opacity effects in nearby dusty
galaxies (e.g.\ \cite{ber97}; \cite{gor97}; \cite{kuc98};
\cite{smi98}).  The principal results from the current work are 
an extensive set of tables, accessible in the on-line version of this 
issue of The Astrophysical Journal or available by contacting the authors.
We will present the details of the models in the following Section 2; 
selected results will be summarized in graphical form in Section 3; 
we will discuss the implications of these results for currently ongoing 
galaxy studies in Section 4, to be followed by a summary in Section 5.

\section{Model}

In order to model the dust radiative transfer in complex systems such
as galaxies, we use a radiative transfer model (named DIRTY) based on
Monte Carlo techniques (\cite{wit77}).  Use of such a model allows for
arbitrary distributions of both the emitting sources (stars) and
absorbing and scattering material (dust).  The full description of
DIRTY can be found in Gordon et al.\ (1999b).  The inputs to DIRTY are
the stellar distribution, the dust distribution, and the dust grain
properties.  With Monte Carlo techniques, individual photons are
emitted isotropically from stars according to the stellar distribution
and followed through a dust distribution.  Each photon's interaction
with the dust is parameterized by the dust optical depth ($\tau$),
albedo, and scattering phase function.  The optical depth determines
where the photon interacts, the albedo gives the probability the
photon is scattered from a dust grain, and the scattering phase
function gives the angle at which the photon scatters.

\subsection{Dust and Stellar Distributions}

As was done in \cite{wit92b}, we have chosen to model spherical
galactic environments instead of entire galaxies.  A galactic
environment describes the distribution of the stars and gas.  
In this paper we concentrate on three galactic environments: CLOUDY,
DUSTY, and SHELL.  These three geometries span the range of star/dust
distributions and are pictorially represented in
Figure~\ref{fig_geometries}.  The CLOUDY geometry has dust extending
to 0.69 of the system radius and stars extending to the system radius.
The DUSTY geometry has both dust and stars extending to the system
radius.  This geometry represents a uniform mixture of stars and dust.
The SHELL geometry has stars extending to 0.3 of the system radius and
dust extending from 0.3 to 1 of the system radius.  This geometry was
named nuclei in \cite{wit92b}.

These three geometries characterize the {\it global} star and dust
distribution.  The {\it local} distribution of the dust can be either
homogeneous or clumpy.  For this paper, we define the clumpy
distribution (using the definitions of \cite{wit96}) to have a filling
factor ({\it ff}) of 0.15, a low-density to high-density ratio
($k_2/k_1$) of 0.01, and a system size characterized by a cube divided
into $30^3$ cubical bins ($N = 30$). A forthcoming third paper in this
series will present the radiative transfer through doubly exponential
disk/bulge galaxies with clumpy ISM.

\vspace*{0.05in}
\epsscale{0.5}
\begin{center}
\plotone{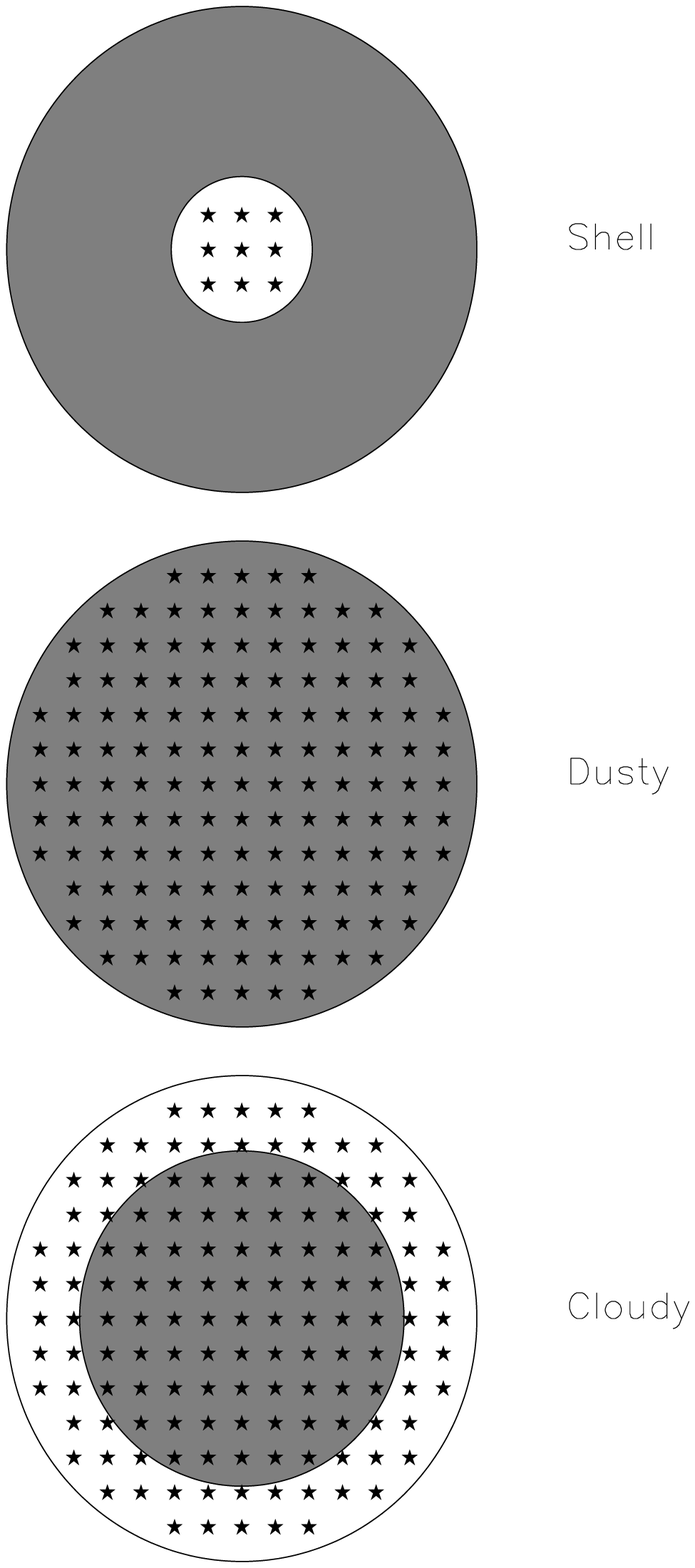}
\end{center}
\figcaption{Pictorial representation of the three geometries used in this
paper (SHELL, DUSTY, and CLOUDY).  The star symbols trace the stellar
distribution and the grey solid regions give the dust distribution.
\label{fig_geometries}} 
\vspace*{0.05in}

\subsection{Dust Grain Characteristics}

We have chosen to compute models using both Milky Way (MW) and Small
Magellanic Cloud (SMC) dust.  By using these two different types of
dust, we are spanning the range of {\it known} dust extinction curves.
For Monte Carlo radiative transfer, we need to know the dependence of
the dust's optical depth ($\tau$), albedo ($a$), and scattering phase
function on wavelength.  In this work, we used the
\cite{hen41} phase function which is parameterized by the
phase function weighted average of the cosine of the scattering angle
($g = \left< cos \theta \right>$).

\begin{figure*}[tbp]
\epsscale{1.0}
\plottwo{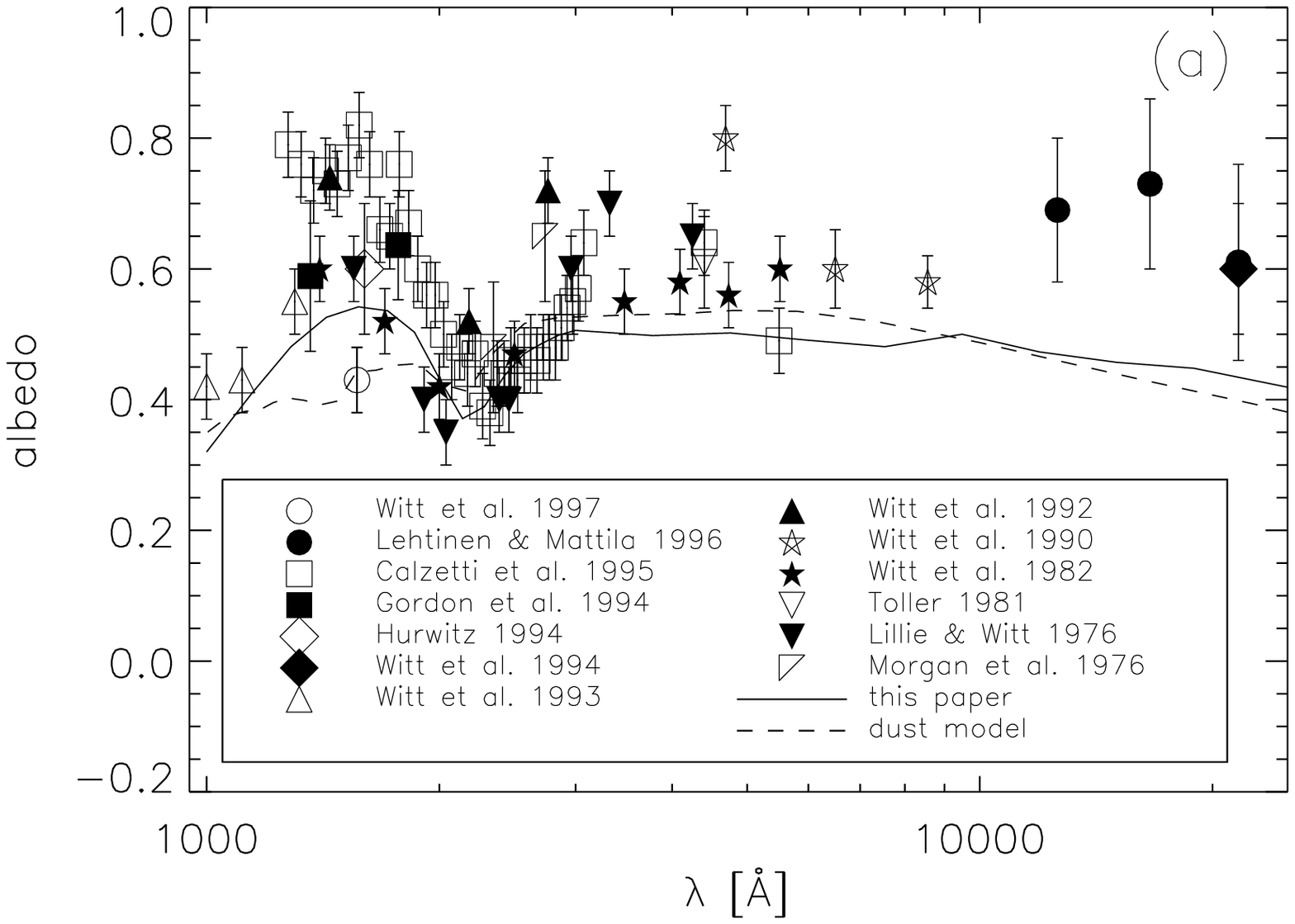}{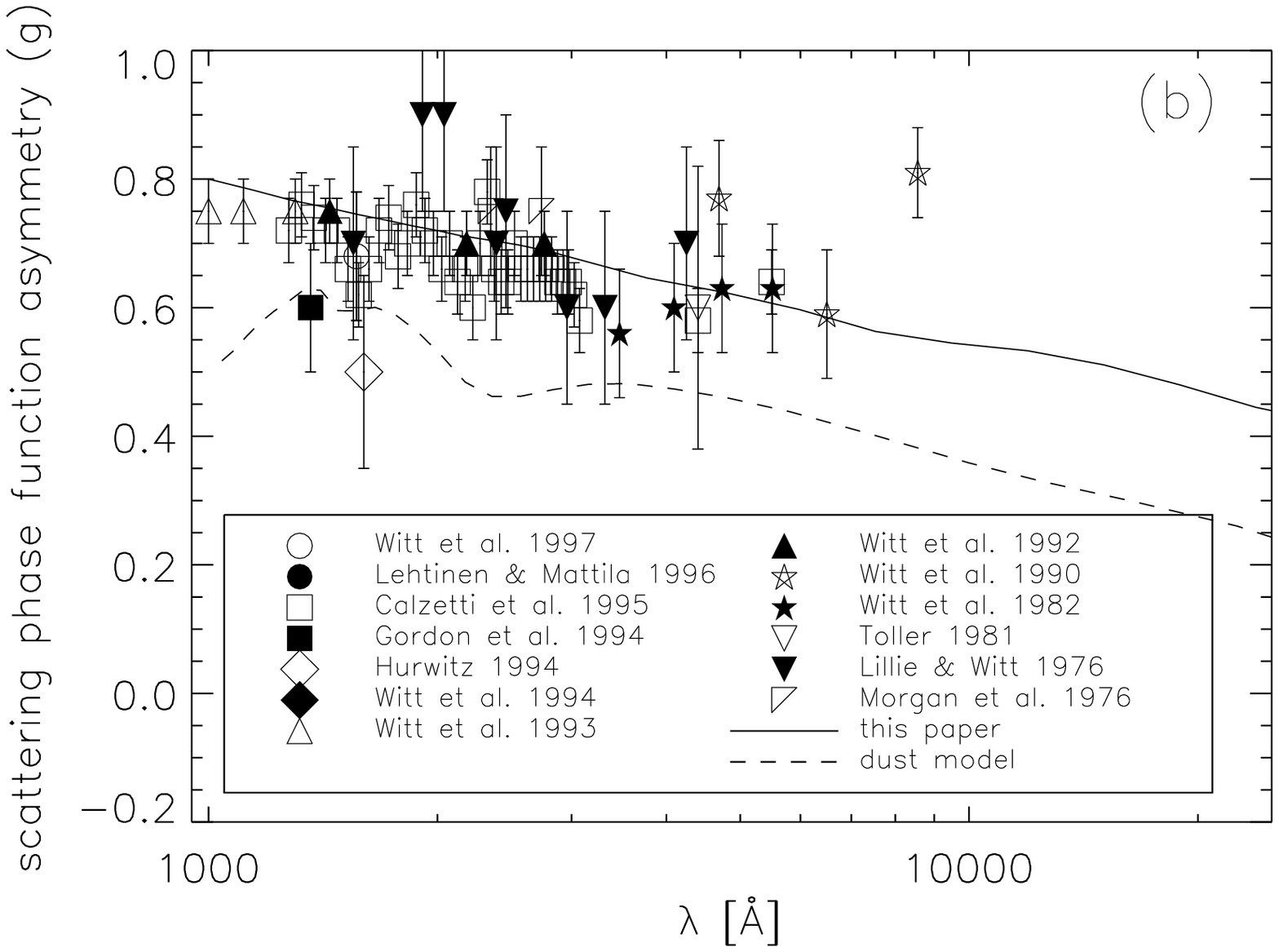}
\caption{The albedo ($a$) and scattering phase function asymmetry
($g$) values are plotted as a function of wavelength in (a) and (b),
respectively.  This plot includes values from empirical determinations
from the literature, a dust grain model (Kim et al.\ 1994), and
numbers adopted for this paper. \label{fig_dust_a_g}}
\end{figure*}

\begin{figure*}[tbp]
\epsscale{1.0}
\plottwo{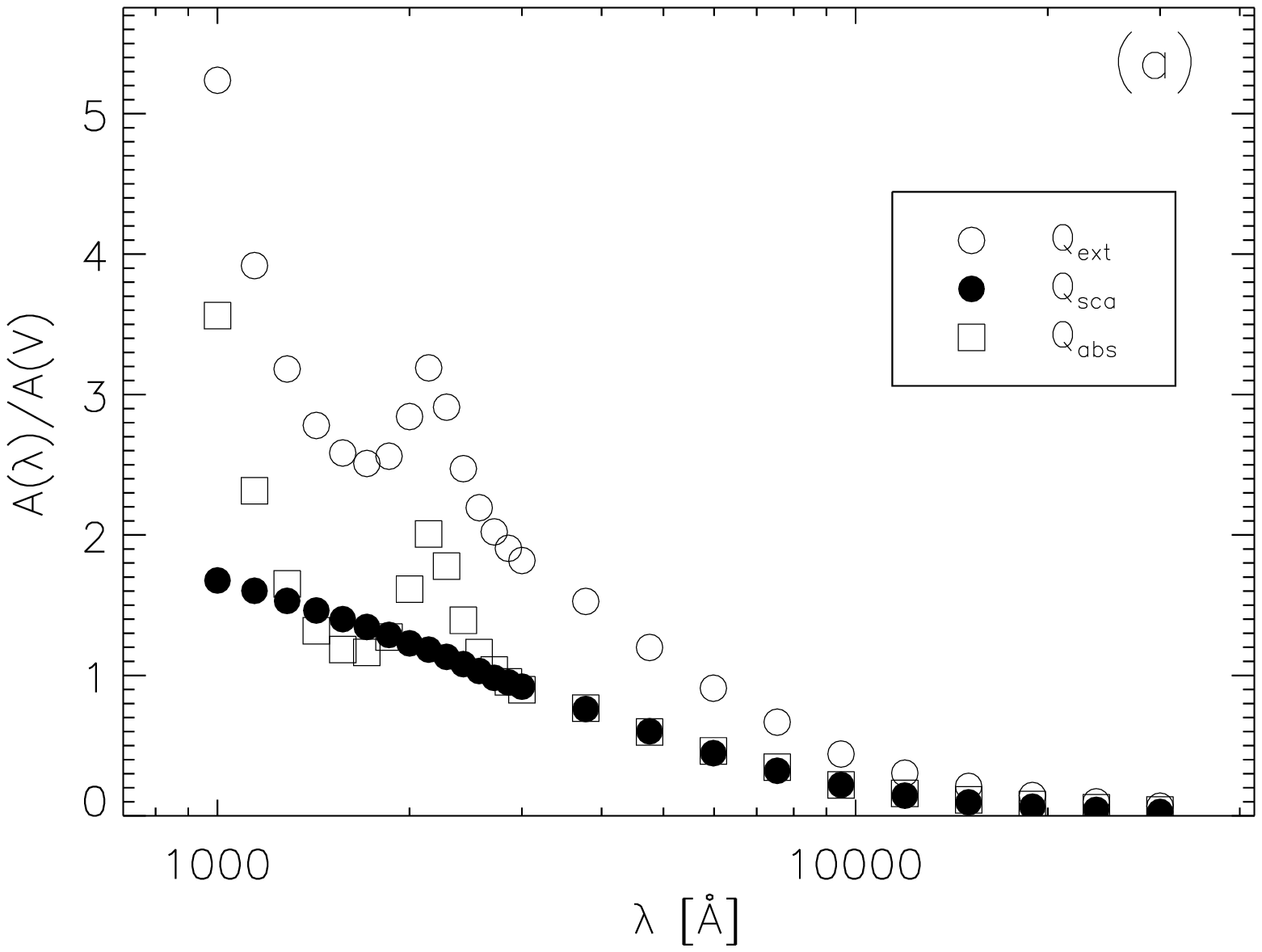}{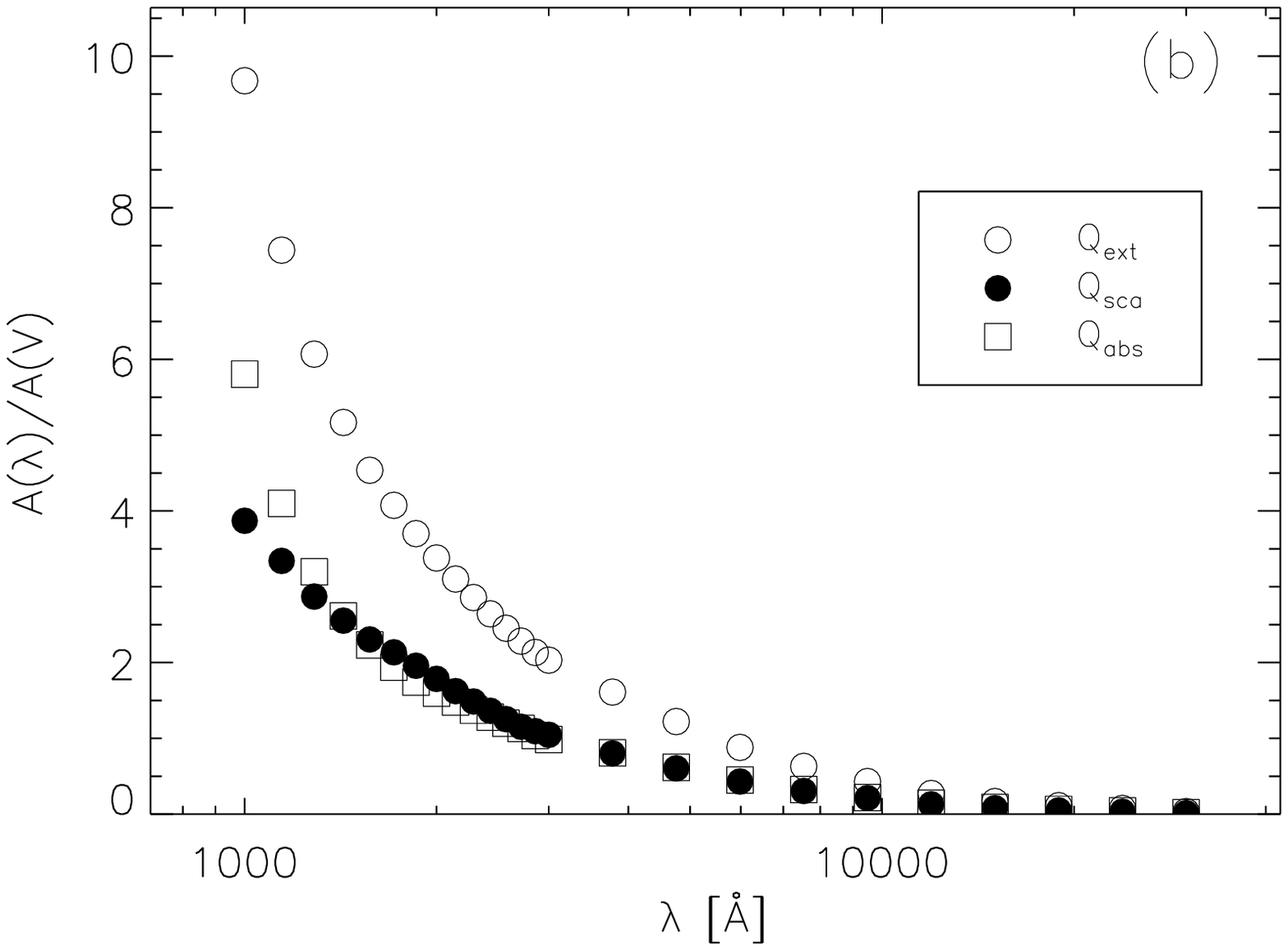}
\caption{The various Q values for the MW and SMC are plotted in (a)
and (b), respectively. \label{fig_dust_Q}}
\end{figure*}

For the MW, we used dust which is associated with the diffuse
interstellar medium and has $R_V = A_V/E(B-V) = 3.1$.  The wavelength
dependence of $\tau$ was determined from the parameterization of Milky
Way extinction curves by Cardelli, Clayton, \& Mathis (1991, hereafter
\cite{car89}).  The $a$ and $g$ values were determined by comparing
empirical determinations of these numbers to a dust grain model which
reproduces the $R_V = 3.1$ extinction curve (\cite{kim94}).
Figure~\ref{fig_dust_a_g}a shows the empirical determinations, dust
grain model values, and the numbers adopted in this paper.  In
determining the albedo values to use for this paper, we visually
averaged the empirical determinations and then shifted the average
to agree with the dust grain model values.  The shift was done because most
of the empirical determinations were done for reflection nebulae which
usually have higher values of $R_V$ than 3.1.  The dust grain model
predicts decreasing albedo's for decreasing $R_V$'s.  This is also
consistent with recent work on the ultraviolet diffuse galactic light
(\cite{wit97}).  When we averaged the albedo values, we also made sure
that $Q_{\rm sca}$ ($= aQ_{\rm ext}$) was smooth across the 2175~\AA\
bump as it is an absorption feature (\cite{cal95}).
Figure~\ref{fig_dust_Q}a plots $Q_{\rm ext}$, $Q_{\rm sca}$, and
$Q_{\rm abs}$ for the MW albedo values used in this paper.  The $g$
values shown in Fig. 2b were taken as averages of the empirical 
determinations, as the
$g$ values from the ultraviolet diffuse galactic light work are not
significantly different from those of work on reflection nebulae.
Also, the dust grain model still has problems in that it predicts the
2175~\AA\ bump as partly a scattering feature, contrary to
observational evidence.  The $\tau$, $a$, and $g$ values adopted for
MW dust in this paper are given in Table~\ref{tab_dust_prop}.

For the SMC, we used dust which is associated with the star forming
Bar of the SMC (\cite{gor98}).  This dust produces extinction curves
which are quite different from dust extinction curves in the MW
(\cite{car89}) or the Large Magellanic Cloud (\cite{mis99}).  It lacks
a 2175~\AA\ bump and is roughly linear with $\lambda^{-1}$.  This type
of dust seems to be associated with intense star forming regions
(\cite{gor97}; \cite{gor98}; \cite{mis99}).  The $\tau$ values were
taken from an average of the three known extinction curves in the SMC
Bar (\cite{gor98}).  There is no empirical work on the $a$ and $g$
values in the SMC.  As there are also still problems with dust grain
models reproducing the MW $a$ and $g$ values, we have chosen to start
with the MW $a$ values and then modify them to reflect the lack of a
2175~\AA\ bump.  This time, we required that the $Q_{\rm sca}$ and
$Q_{\rm abs}$ values did not show any features across the 2175~\AA\
region.  Figure~\ref{fig_dust_Q}b displays $Q_{\rm ext}$, $Q_{\rm
sca}$, and $Q_{\rm abs}$ for the SMC albedo values used in this paper.
The $\tau$, $a$, and $g$ values adopted for SMC dust in this paper are
given in Table~\ref{tab_dust_prop}.

\begin{deluxetable}{ccccccc}
\tablewidth{0pt}
\tablecaption{Dust Physical Properties \label{tab_dust_prop}}
\tablehead{ & \multicolumn{3}{c}{MW} & 
            \multicolumn{3}{c}{SMC} \\
           \colhead{$\lambda$ [\AA]} & \colhead{$\tau/\tau_V$} & 
             \colhead{a} & \colhead{g} & \colhead{$\tau/\tau_V$} & 
             \colhead{a} & \colhead{g}} 
\startdata
 1000.0 & 5.238 & 0.320 & 0.800 & 9.675 & 0.400 & 0.800 \nl
 1142.9 & 3.918 & 0.409 & 0.783 & 7.440 & 0.449 & 0.783 \nl
 1285.7 & 3.182 & 0.481 & 0.767 & 6.068 & 0.473 & 0.767 \nl
 1428.6 & 2.780 & 0.526 & 0.756 & 5.167 & 0.494 & 0.756 \nl
 1571.4 & 2.584 & 0.542 & 0.745 & 4.536 & 0.508 & 0.745 \nl
 1714.3 & 2.509 & 0.536 & 0.736 & 4.074 & 0.524 & 0.736 \nl
 1857.1 & 2.561 & 0.503 & 0.727 & 3.700 & 0.529 & 0.727 \nl
 2000.0 & 2.843 & 0.432 & 0.720 & 3.379 & 0.528 & 0.720 \nl
 2142.9 & 3.190 & 0.371 & 0.712 & 3.101 & 0.523 & 0.712 \nl
 2285.7 & 2.910 & 0.389 & 0.707 & 2.857 & 0.520 & 0.707 \nl
 2428.6 & 2.472 & 0.437 & 0.702 & 2.642 & 0.516 & 0.702 \nl
 2571.4 & 2.194 & 0.470 & 0.697 & 2.452 & 0.511 & 0.697 \nl
 2714.3 & 2.022 & 0.486 & 0.691 & 2.282 & 0.505 & 0.691 \nl
 2857.1 & 1.905 & 0.499 & 0.685 & 2.133 & 0.513 & 0.685 \nl
 3000.0 & 1.818 & 0.506 & 0.678 & 2.031 & 0.515 & 0.678 \nl
 3776.8 & 1.527 & 0.498 & 0.646 & 1.610 & 0.498 & 0.646 \nl
 4754.7 & 1.199 & 0.502 & 0.624 & 1.221 & 0.494 & 0.624 \nl
 5985.8 & 0.909 & 0.491 & 0.597 & 0.880 & 0.489 & 0.597 \nl
 7535.8 & 0.667 & 0.481 & 0.563 & 0.630 & 0.484 & 0.563 \nl
 9487.0 & 0.440 & 0.500 & 0.545 & 0.430 & 0.493 & 0.545 \nl
11943.5 & 0.304 & 0.473 & 0.533 & 0.272 & 0.475 & 0.533 \nl
15036.0 & 0.210 & 0.457 & 0.511 & 0.166 & 0.465 & 0.511 \nl
18929.2 & 0.145 & 0.448 & 0.480 & 0.111 & 0.439 & 0.480 \nl
23830.6 & 0.100 & 0.424 & 0.445 & 0.075 & 0.417 & 0.445 \nl
30001.0 & 0.069 & 0.400 & 0.420 & 0.033 & 0.400 & 0.420 \nl
\enddata
\end{deluxetable}

\section{Results}

\subsection{Structure of the Data Tables}

\begin{figure*}[tbp]
\epsscale{1.0}
\plottwo{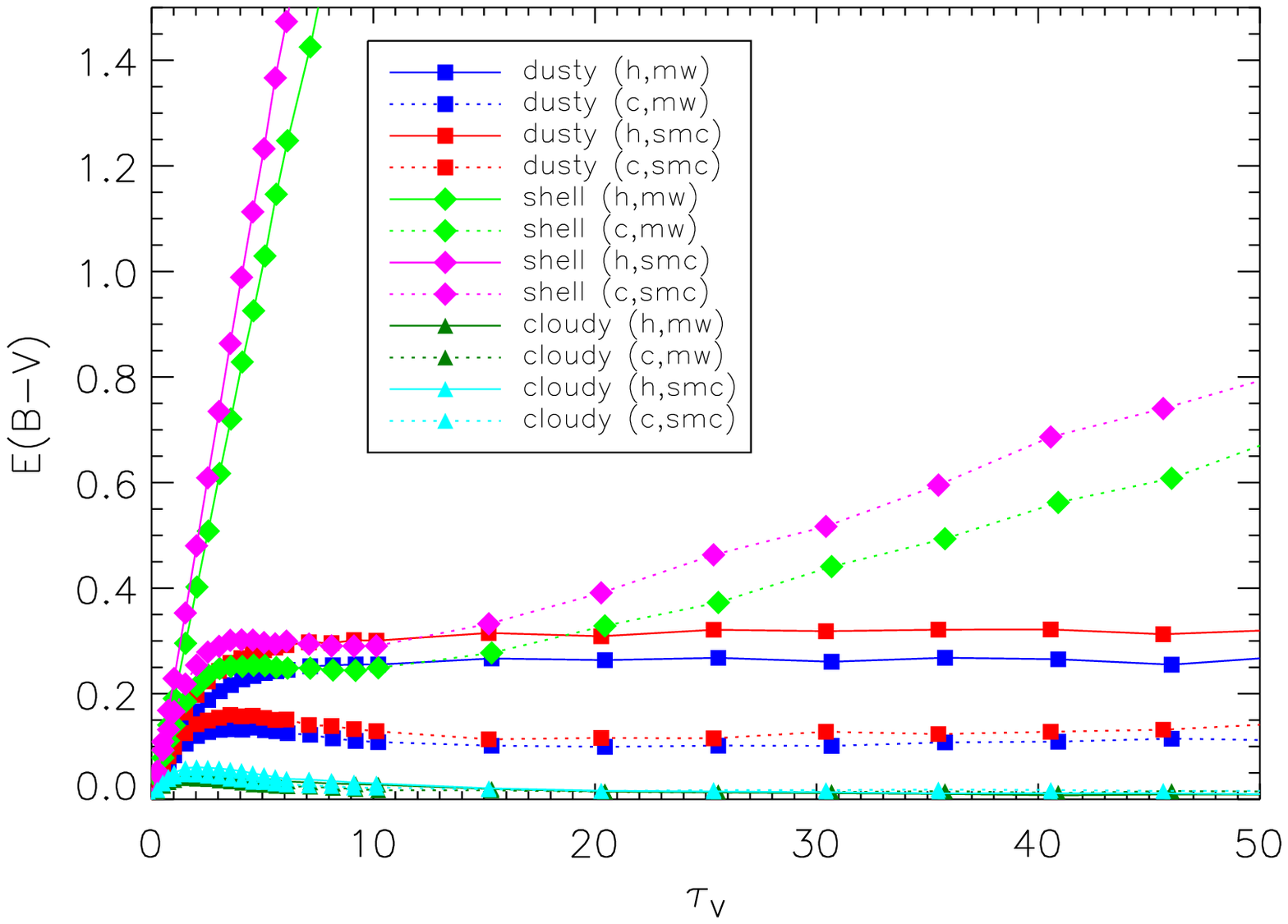}{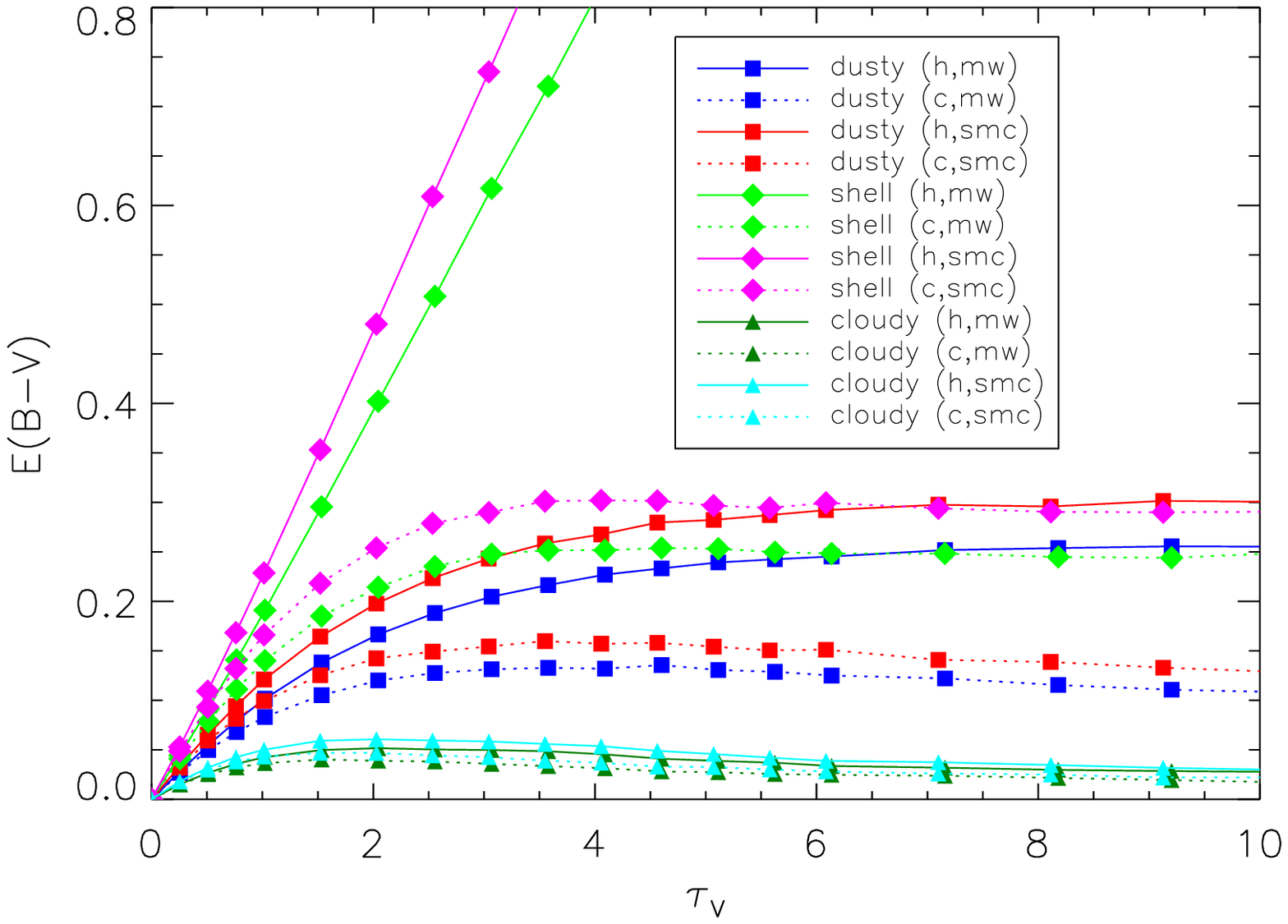}
\caption{The traditional reddening signature of dust [$E(B-V)$] is
plotted versus the input $\tau_V$ (proportional to dust mass) in (a).
In (b), a blowup of this plot is shown to illustrate the behavior of
E(B-V) for $\tau_V$ values $< 10$. \label{fig_ebv_tauV}}
\end{figure*}

The results of our radiative transfer calculations are contained in
Tables 2, 3, and 4, accessible in full in the online edition of this issue
or available directly from the authors.  Table 2
contains the data for the DUSTY galactic environment, Table 3 those
for the SHELL galactic environment, and Table 4 those for the CLOUDY
environment.  Only the beginning of Table 2 is shown in this paper in
order to illustrate the type of data available in the full set of
tables. Each model is identified by its geometry (DUSTY, SHELL,
CLOUDY), the intrinsic opacity function (MW, SMC), the radial
extinction optical depth from the center to the edge of the dust
environment at V ($\tau_V =$ 0.25, 0.5, ...,50), assuming a constant
density homogeneous distribution, and the type of structure
(homogeneous, clumpy) in the dust region. Each individual model
contains calculations for 25 wavelengths, ranging from 1000~\AA\ to
30000~\AA, shown in column 1. Column 2 identifies the optical depths
for extinction at each wavelength, derived from the applicable
intrinsic opacity function. Columns 3, 4, 5, and 6 list, respectively,
the attenuation optical depth ($\tau_{\rm att}$), the scattered flux 
fraction, the direct attenuated stellar flux fraction, and the total 
escaping flux fraction, containing both attenuated stellar as well as
scattered radiation, for the homogeneous case. Columns
7 to 10 repeat this information for the equivalent clumpy case. All
fluxes are referred to unity at each wavelength, with unit flux
corresponding to no attenuation, and in cases with clumpy dust
distributions, they are averages over all directions.

For a given geometry, the value of the V-band optical depth may be
regarded as a measure of the dust mass present in that geometry.  The
attenuation optical depth (column 3 and 7) is defined as
\begin{equation}
\tau_{\rm att} = -\ln({\rm f(esc)})
\end{equation}
and thus measures the absorption occurring in the system. This is a
rather useful measurement since in galactic environments viewed from
afar both the attenuated stellar flux as well as the
scattered light are combined in the escaping total flux observed. Note that
this definition of the attenuation optical depth differs from that of
the effective optical depth ($\tau_{\rm eff}$) defined in
\cite{wit96}, which refers to the attenuated, non-scattered stellar
flux only.  The $\tau_{\rm att}$ values are always smaller than the
radial optical depth listed in column 2, because the dust albedo is
always non-zero and the stars are distributed throughout extended
volumes. Also, the attenuation optical depth for the clumpy cases
never exceeds that of the corresponding homogeneous cases
(\cite{wit96}). It needs to be reemphasized that the clumpy case treated here
is for one specific set of clumpiness parameters, namely a filling factor 
ff = 0.15, an interclump medium to clump density ratio $k_2/k_1$ = 0.01,
and a system size divided into $30^3$ cubical bins. The variations of 
transfer characteristics expected when any of these parameters are changed
within reasonable ranges were explored in \cite{wit96}, and they are 
briefly discussed in Section 3.5.

\begin{deluxetable}{rc|cccc|cccc}
\tablewidth{0pt}
\tablecaption{Dusty Galactic Environment \label{tab_dusty_sge}}
\tablehead{\colhead{$\lambda$} & \colhead{$\tau$} & 
           \colhead{$\tau_{\rm att}$} & \colhead{f(sca)} & 
           \colhead{f(dir)} & \colhead{f(esc)} &
           \colhead{$\tau_{\rm att}$} & \colhead{f(sca)} & 
           \colhead{f(dir)} & \colhead{f(esc)}}
\startdata
\multicolumn{2}{c}{dusty,MW} & \multicolumn{4}{c}{$\tau_V =  0.25$, homogeneous} & \multicolumn{4}{c}{$\tau_V =  0.25$, clumpy} \nl \tableline
  1000 &   1.309 &   0.589 & 1.07E-01 & 4.48E-01 & 5.55E-01 &   0.517 & 8.84E-02 & 5.08E-01 & 5.96E-01 \nl
  1142 &   0.979 &   0.402 & 1.37E-01 & 5.32E-01 & 6.69E-01 &   0.366 & 1.16E-01 & 5.77E-01 & 6.93E-01 \nl
  1285 &   0.795 &   0.295 & 1.54E-01 & 5.90E-01 & 7.45E-01 &   0.274 & 1.34E-01 & 6.26E-01 & 7.60E-01 \nl
  1428 &   0.695 &   0.238 & 1.62E-01 & 6.26E-01 & 7.88E-01 &   0.225 & 1.42E-01 & 6.56E-01 & 7.99E-01 \nl
  1571 &   0.646 &   0.215 & 1.61E-01 & 6.45E-01 & 8.06E-01 &   0.204 & 1.43E-01 & 6.72E-01 & 8.15E-01 \nl
  1714 &   0.627 &   0.212 & 1.56E-01 & 6.53E-01 & 8.09E-01 &   0.201 & 1.39E-01 & 6.78E-01 & 8.17E-01 \nl
  1857 &   0.640 &   0.230 & 1.47E-01 & 6.48E-01 & 7.94E-01 &   0.218 & 1.30E-01 & 6.74E-01 & 8.04E-01 \nl
  2000 &   0.711 &   0.289 & 1.29E-01 & 6.20E-01 & 7.49E-01 &   0.270 & 1.13E-01 & 6.51E-01 & 7.64E-01 \nl
  2142 &   0.798 &   0.353 & 1.13E-01 & 5.90E-01 & 7.03E-01 &   0.325 & 9.76E-02 & 6.25E-01 & 7.23E-01 \nl
  2285 &   0.728 &   0.315 & 1.15E-01 & 6.14E-01 & 7.30E-01 &   0.293 & 1.00E-01 & 6.46E-01 & 7.46E-01 \nl
  2428 &   0.618 &   0.251 & 1.22E-01 & 6.56E-01 & 7.78E-01 &   0.236 & 1.08E-01 & 6.82E-01 & 7.90E-01 \nl
  2571 &   0.549 &   0.211 & 1.24E-01 & 6.86E-01 & 8.10E-01 &   0.200 & 1.11E-01 & 7.07E-01 & 8.18E-01 \nl
  2714 &   0.506 &   0.190 & 1.23E-01 & 7.04E-01 & 8.27E-01 &   0.181 & 1.11E-01 & 7.24E-01 & 8.35E-01 \nl
  2857 &   0.476 &   0.175 & 1.22E-01 & 7.17E-01 & 8.39E-01 &   0.167 & 1.10E-01 & 7.36E-01 & 8.46E-01 \nl
  3000 &   0.454 &   0.165 & 1.20E-01 & 7.28E-01 & 8.48E-01 &   0.158 & 1.09E-01 & 7.45E-01 & 8.54E-01 \nl
  3776 &   0.382 &   0.141 & 1.05E-01 & 7.63E-01 & 8.68E-01 &   0.136 & 9.61E-02 & 7.77E-01 & 8.73E-01 \nl
  4754 &   0.300 &   0.111 & 8.85E-02 & 8.07E-01 & 8.95E-01 &   0.107 & 8.24E-02 & 8.16E-01 & 8.99E-01 \nl
  5985 &   0.227 &   0.086 & 6.94E-02 & 8.48E-01 & 9.18E-01 &   0.084 & 6.56E-02 & 8.54E-01 & 9.20E-01 \nl
  7535 &   0.167 &   0.065 & 5.24E-02 & 8.85E-01 & 9.37E-01 &   0.063 & 5.00E-02 & 8.89E-01 & 9.39E-01 \nl
  9487 &   0.110 &   0.041 & 3.77E-02 & 9.22E-01 & 9.60E-01 &   0.040 & 3.65E-02 & 9.24E-01 & 9.60E-01 \nl
 11943 &   0.076 &   0.030 & 2.53E-02 & 9.45E-01 & 9.70E-01 &   0.029 & 2.47E-02 & 9.46E-01 & 9.71E-01 \nl
 15036 &   0.052 &   0.021 & 1.72E-02 & 9.62E-01 & 9.79E-01 &   0.021 & 1.69E-02 & 9.62E-01 & 9.79E-01 \nl
 18929 &   0.036 &   0.015 & 1.18E-02 & 9.73E-01 & 9.85E-01 &   0.015 & 1.16E-02 & 9.74E-01 & 9.86E-01 \nl
 23830 &   0.025 &   0.011 & 7.80E-03 & 9.82E-01 & 9.89E-01 &   0.011 & 7.68E-03 & 9.82E-01 & 9.89E-01 \nl
 30001 &   0.017 &   0.008 & 5.11E-03 & 9.87E-01 & 9.92E-01 &   0.008 & 5.03E-03 & 9.87E-01 & 9.92E-01 \nl
\tableline
\multicolumn{2}{c}{dusty,SMC} & \multicolumn{4}{c}{$\tau_V =  0.25$, homogeneous} & \multicolumn{4}{c}{$\tau_V =  0.25$, clumpy} \nl \tableline
  1000 &   2.419 &   0.887 & 1.29E-01 & 2.83E-01 & 4.12E-01 &   0.733 & 1.04E-01 & 3.76E-01 & 4.80E-01 \nl
  1142 &   1.860 &   0.671 & 1.62E-01 & 3.50E-01 & 5.11E-01 &   0.578 & 1.32E-01 & 4.29E-01 & 5.61E-01 \nl
  1285 &   1.517 &   0.543 & 1.75E-01 & 4.06E-01 & 5.81E-01 &   0.481 & 1.45E-01 & 4.74E-01 & 6.18E-01 \nl
  1428 &   1.292 &   0.455 & 1.83E-01 & 4.51E-01 & 6.35E-01 &   0.409 & 1.53E-01 & 5.11E-01 & 6.64E-01 \nl
  1571 &   1.134 &   0.393 & 1.85E-01 & 4.89E-01 & 6.75E-01 &   0.359 & 1.57E-01 & 5.42E-01 & 6.98E-01 \nl
  1714 &   1.018 &   0.345 & 1.87E-01 & 5.21E-01 & 7.08E-01 &   0.318 & 1.60E-01 & 5.68E-01 & 7.28E-01 \nl
  1857 &   0.925 &   0.313 & 1.83E-01 & 5.48E-01 & 7.32E-01 &   0.290 & 1.58E-01 & 5.91E-01 & 7.49E-01 \nl
  2000 &   0.845 &   0.287 & 1.77E-01 & 5.74E-01 & 7.50E-01 &   0.268 & 1.53E-01 & 6.11E-01 & 7.65E-01 \nl
  2142 &   0.775 &   0.267 & 1.68E-01 & 5.97E-01 & 7.65E-01 &   0.250 & 1.47E-01 & 6.32E-01 & 7.79E-01 \nl
  2285 &   0.714 &   0.249 & 1.61E-01 & 6.19E-01 & 7.80E-01 &   0.234 & 1.41E-01 & 6.50E-01 & 7.91E-01 \nl
  2428 &   0.660 &   0.232 & 1.53E-01 & 6.40E-01 & 7.93E-01 &   0.219 & 1.36E-01 & 6.68E-01 & 8.03E-01 \nl
  2571 &   0.613 &   0.218 & 1.46E-01 & 6.58E-01 & 8.04E-01 &   0.206 & 1.29E-01 & 6.84E-01 & 8.14E-01 \nl
  2714 &   0.571 &   0.206 & 1.38E-01 & 6.76E-01 & 8.14E-01 &   0.196 & 1.23E-01 & 6.99E-01 & 8.22E-01 \nl
  2857 &   0.533 &   0.190 & 1.35E-01 & 6.92E-01 & 8.27E-01 &   0.181 & 1.21E-01 & 7.13E-01 & 8.35E-01 \nl
  3000 &   0.508 &   0.181 & 1.31E-01 & 7.03E-01 & 8.35E-01 &   0.173 & 1.19E-01 & 7.23E-01 & 8.41E-01 \nl
  3776 &   0.403 &   0.149 & 1.09E-01 & 7.53E-01 & 8.62E-01 &   0.143 & 9.93E-02 & 7.68E-01 & 8.67E-01 \nl
  4754 &   0.305 &   0.115 & 8.82E-02 & 8.04E-01 & 8.92E-01 &   0.111 & 8.19E-02 & 8.13E-01 & 8.95E-01 \nl
  5985 &   0.220 &   0.084 & 6.73E-02 & 8.52E-01 & 9.20E-01 &   0.081 & 6.36E-02 & 8.58E-01 & 9.22E-01 \nl
  7535 &   0.157 &   0.061 & 5.02E-02 & 8.91E-01 & 9.41E-01 &   0.059 & 4.81E-02 & 8.94E-01 & 9.42E-01 \nl
  9487 &   0.108 &   0.041 & 3.64E-02 & 9.24E-01 & 9.60E-01 &   0.040 & 3.53E-02 & 9.26E-01 & 9.61E-01 \nl
 11943 &   0.068 &   0.027 & 2.29E-02 & 9.51E-01 & 9.74E-01 &   0.026 & 2.24E-02 & 9.52E-01 & 9.74E-01 \nl
 15036 &   0.041 &   0.017 & 1.40E-02 & 9.69E-01 & 9.83E-01 &   0.016 & 1.37E-02 & 9.70E-01 & 9.84E-01 \nl
 18929 &   0.028 &   0.012 & 8.95E-03 & 9.80E-01 & 9.89E-01 &   0.012 & 8.79E-03 & 9.80E-01 & 9.89E-01 \nl
 23830 &   0.019 &   0.008 & 5.78E-03 & 9.86E-01 & 9.92E-01 &   0.008 & 5.70E-03 & 9.86E-01 & 9.92E-01 \nl
 30001 &   0.008 &   0.003 & 2.47E-03 & 9.94E-01 & 9.97E-01 &   0.003 & 2.43E-03 & 9.94E-01 & 9.97E-01 \nl
\enddata
\end{deluxetable}

\subsection{Reddening Effects}

Reddening of a spectral energy distribution compared to that expected
from spectroscopic indicators is usually taken as the first sign of
the presence of dust in a galactic system. Color-color diagrams
derived from multi-color photometry of galaxies frequently exhibit
reddening {\em arrows} which purport to show the direction in which
data points move in the color-color plane in response to the presence
of a given column density of dust.  Implicit in such displays is the
assumption that the reddening is proportional to the dust column. It
was shown by \cite{wit92b} that the relationships between reddenings
and dust columns is usually quite non-linear and highly dependent upon
details of the geometry and the scattering properties of the dust.
Here we show the additional effects due to the presence of a clumpy
structure in the attenuating medium.

In Fig.~\ref{fig_ebv_tauV}a we show the traditional reddening
parameter $E(B-V)$ as a function of the homogeneous radial optical depth $\tau_V$;
in Fig.~\ref{fig_ebv_tauV}b the congested region near the origin of
Fig.~\ref{fig_ebv_tauV}a is enlarged. Several interesting points can
be made. The most dramatic effect arising from the transition from a
homogeneous structure to a two-phase clumpy structure occurs in the
SHELL geometry. The homogeneous SHELL is similar to a spherical
homogeneous screen, except that flux is scattered back into the beam
in the SHELL geometry.  As expected for a screen, the predicted
reddening for the homogeneous SHELL geometry increases linearly with
increasing dust column density.  The scaling factor is less than that
for a screen due to the blue color of scattered radiation included in
the measurement.  Consequently, the ratios $E(B-V)/\tau_V$ for the
homogeneous SHELL models are 0.20 for the MW case, 0.24 for the SMC
case. This should be compared to the standard ratio of 0.35 for
average Galactic extinction. The transition to a two-phase clumpy dust
structure in the shell leads to a reddening function which quickly
becomes non-linear and saturates near $E(B-V) \sim 0.25$ for MW dust,
near $E(B-V) \sim 0.30$ for SMC dust. This initial saturation is
caused by the fact that individual clumps become optically thick while
the low-density interclump medium initially contributes little to the
reddening. At $\tau_V > 10$, the clumpy SHELL geometry exhibits a
gradual increase in the reddening again, now solely due to the growing
attenuation by the interclump medium. Note the interesting parallel to
the behavior of a curve-of-growth in astronomical spectroscopy, where
the increase of absorption in the core of a line is followed by
saturation and subsequent further growth due to absorption in the
damping wings of the line.

The DUSTY geometry exhibits a permanent saturation at larger optical
depths both in the clumpy and in the homogeneous states, as was shown
already by \cite{wit92b}. The transition to a clumpy structure makes
the saturation occur at roughly half the amount of reddening in
$E(B-V)$ than was found in the homogeneous case. The saturation in the
homogeneous case is a result of the fact that radiation emerges
primarily from the outermost regions within optical depths one to two,
regardless of how much dust is present further inside the system. This
is fundamentally unchanged by adopting a clumpy structure, except that
light now escapes preferentially through low-optical depth
lines-of-sight between clumps, resulting in an overall lower amount of
reddening.

The level of $E(B-V)$ at which realistic dusty structures saturate is
highly significant. It has been noted with great emphasis by Dickinson
(1997), Sawicki and Yee (1998), and Steidel et al.\ (1999), among
others, that the $E(B-V)$ color excesses of Lyman-break galaxies crowd
around values of $E(B-V) \sim 0.2$, with hardly any exhibiting color
excesses greater than 0.35.  This result is fully consistent with our
model results. All models except the homogeneous SHELL models saturate
at reddening values $E(B-V)$ less than 0.3, a value which is breached
by the clumpy SHELL models only for $\tau_V > 15$. This suggests that
the highly idealized homogeneous SHELL model, which is basically the
historical screen model with scattering, does not apply to these
Lyman-break galaxies, as was argued forcefully by \cite{wit92b}
already.  However, the result also means that the actual dust column
densities and the resulting attenuations are quite unconstrained by
these color excess measurements. The fact that these galaxies were
selected on the basis of their rest-frame-UV emission characteristics
does argue, however, for moderate total attenuations.

The saturation of color excesses occurs for higher column densities of
dust when wavelengths longer than the V-band are involved. In
Fig.~\ref{fig_evk_ebv}, this is illustrated in the case of a $E(V-K)$
versus $E(B-V)$ color-color plot. Here, models with increasing optical
depths are plotted as series of connected symbols, starting with
$\tau_V = 0$ at the origin. The separation of symbols from one another
along a given reddening line is a measure of how each color excess
scales with optical depth. Not only is the reddening line not a simple
{\em arrow} (with the exception of the homogeneous SHELL model), the
detailed shape of the reddening line depends on the dust/star geometry
and the structure of the dusty medium. Of particular interest again is
the clumpy SHELL geometry, which shows the initial saturation in
$E(B-V)$ when clumps become optically thick in the V-band, followed by
a saturation in $E(V-K)$ at much higher radial dust column density
when the same clumps reach the optically thick state in the
K-band. Finally, after clumps are sufficiently thick to block light in
all bands, the interclump medium begins to dominate the reddening
process. Where and when this occurs depends on the ratio of densities
in the clump and the interclump medium.  In a real physical system
where the differences between different phases are rather more gradual
than the fixed ratio assumed in our models, the transition shown in
Fig. 5 between the reddening caused by the clumps and that caused by
the interclump medium will also occur in a less abrupt manner.

\vspace*{0.05in}
\epsscale{0.45}
\begin{center}
\plotone{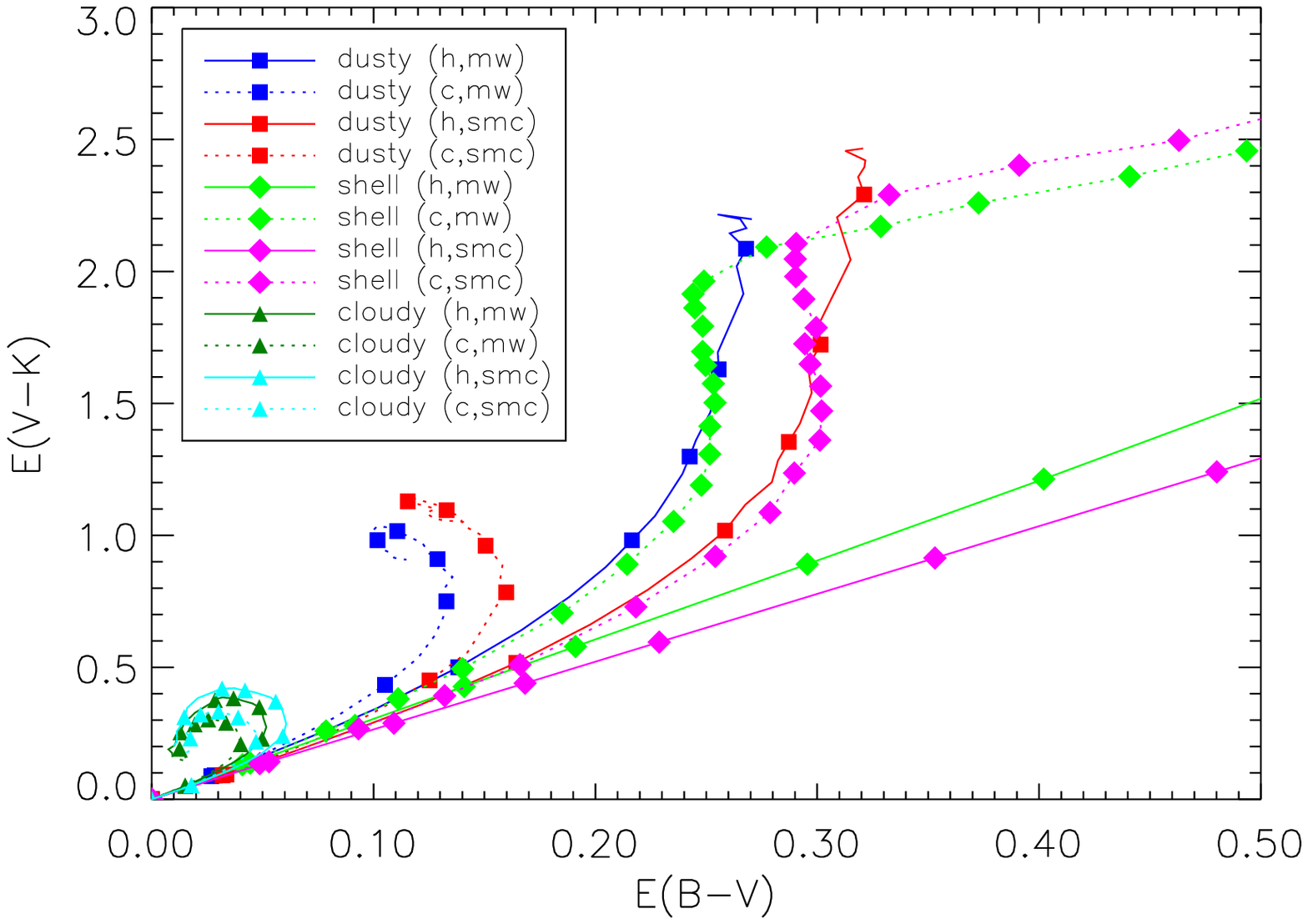}
\end{center}
\figcaption{$E(V-K)$ is plotted versus $E(B-V)$. The location of the
symbols along each line indicated the change in colors as the optical
depth is increased.  For the SHELL geometry, the symbols are plotted
for every model $\tau_V$ point.  For the DUSTY and CLOUDY, the symbols
are plotted for every fourth model $\tau_V$ point.
\label{fig_evk_ebv}}
\vspace*{0.05in}

\subsection{Attenuation Functions}

\begin{figure*}[tbp]
\plottwo{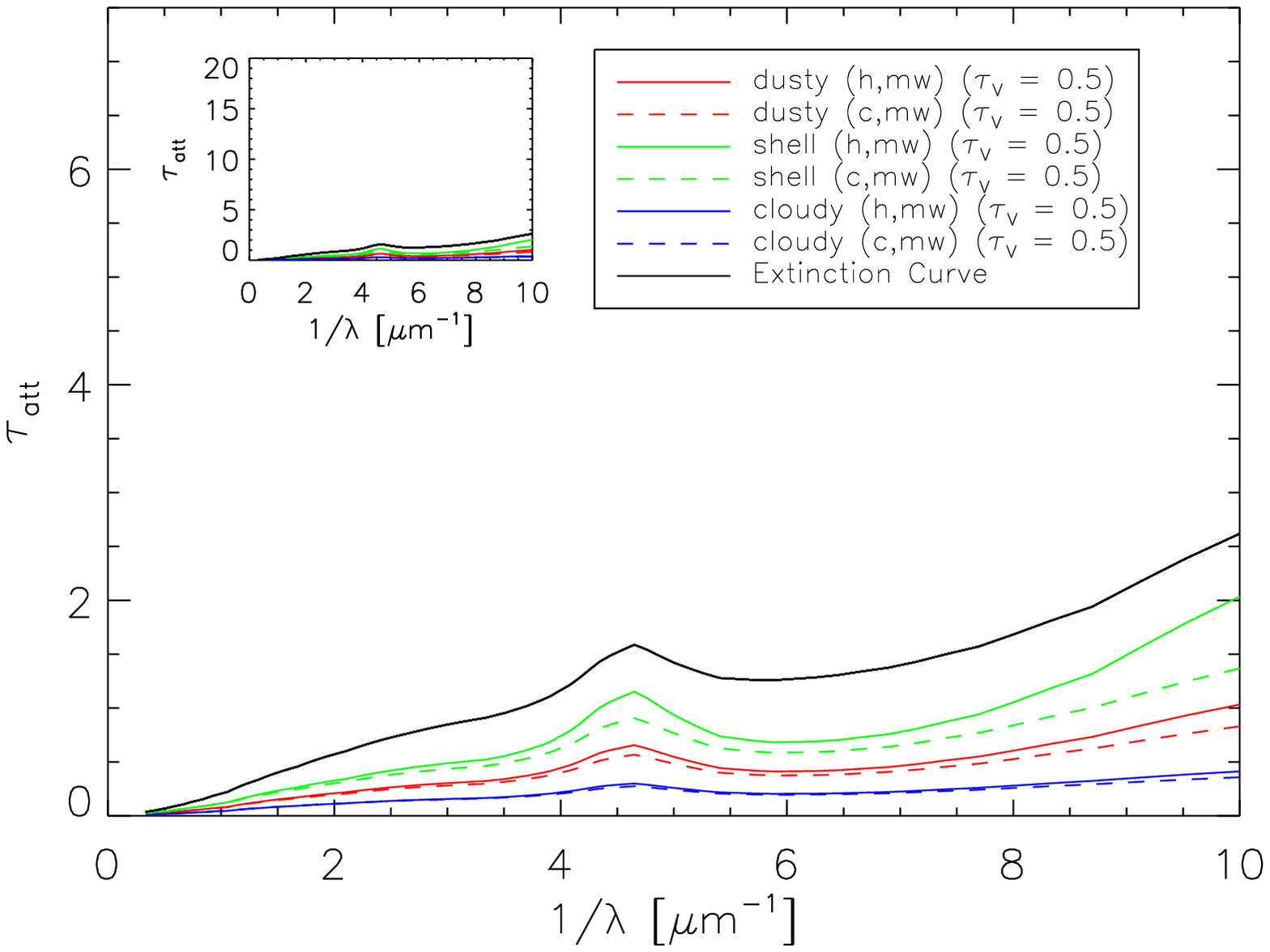}{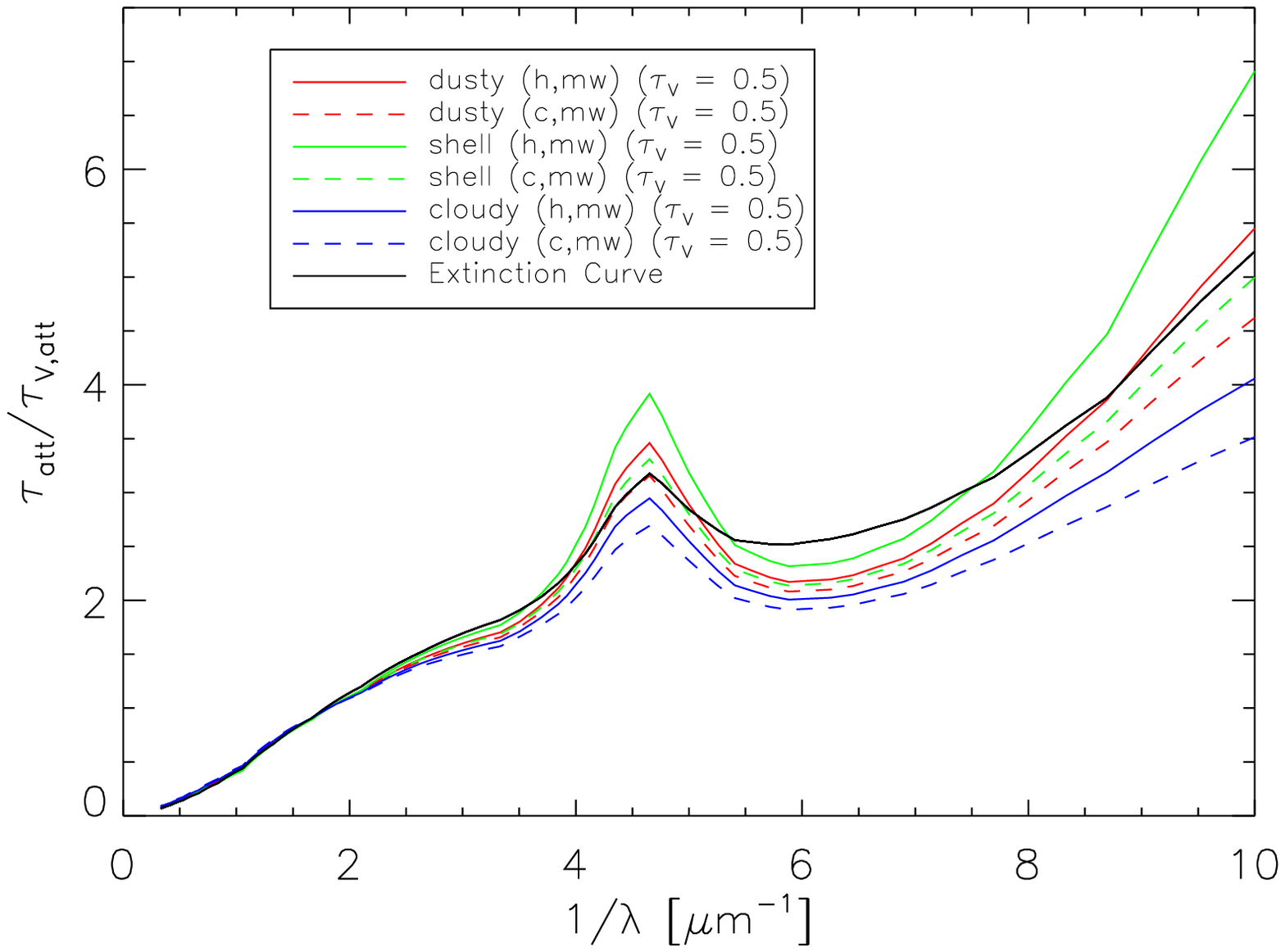} \\
\plottwo{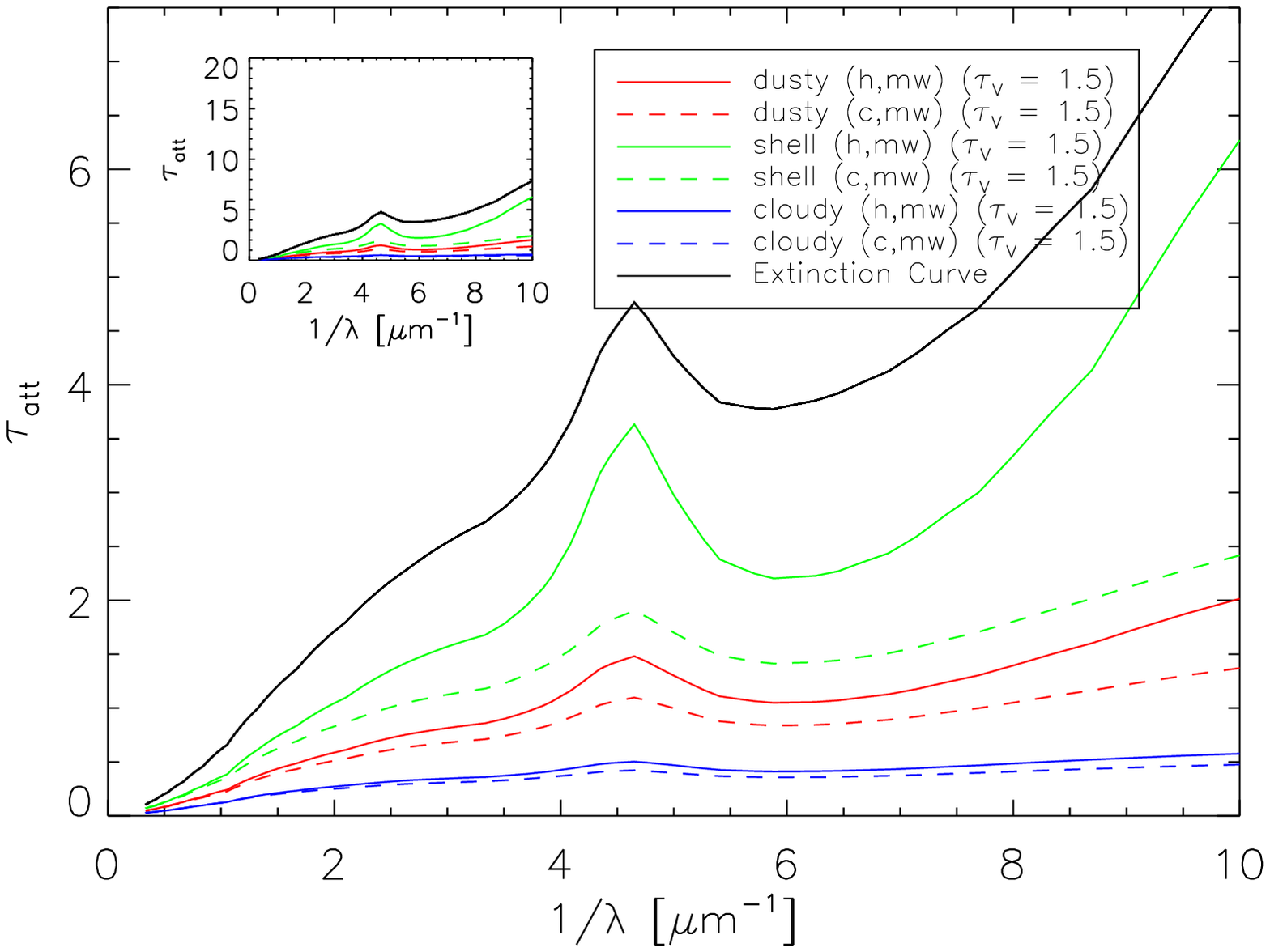}{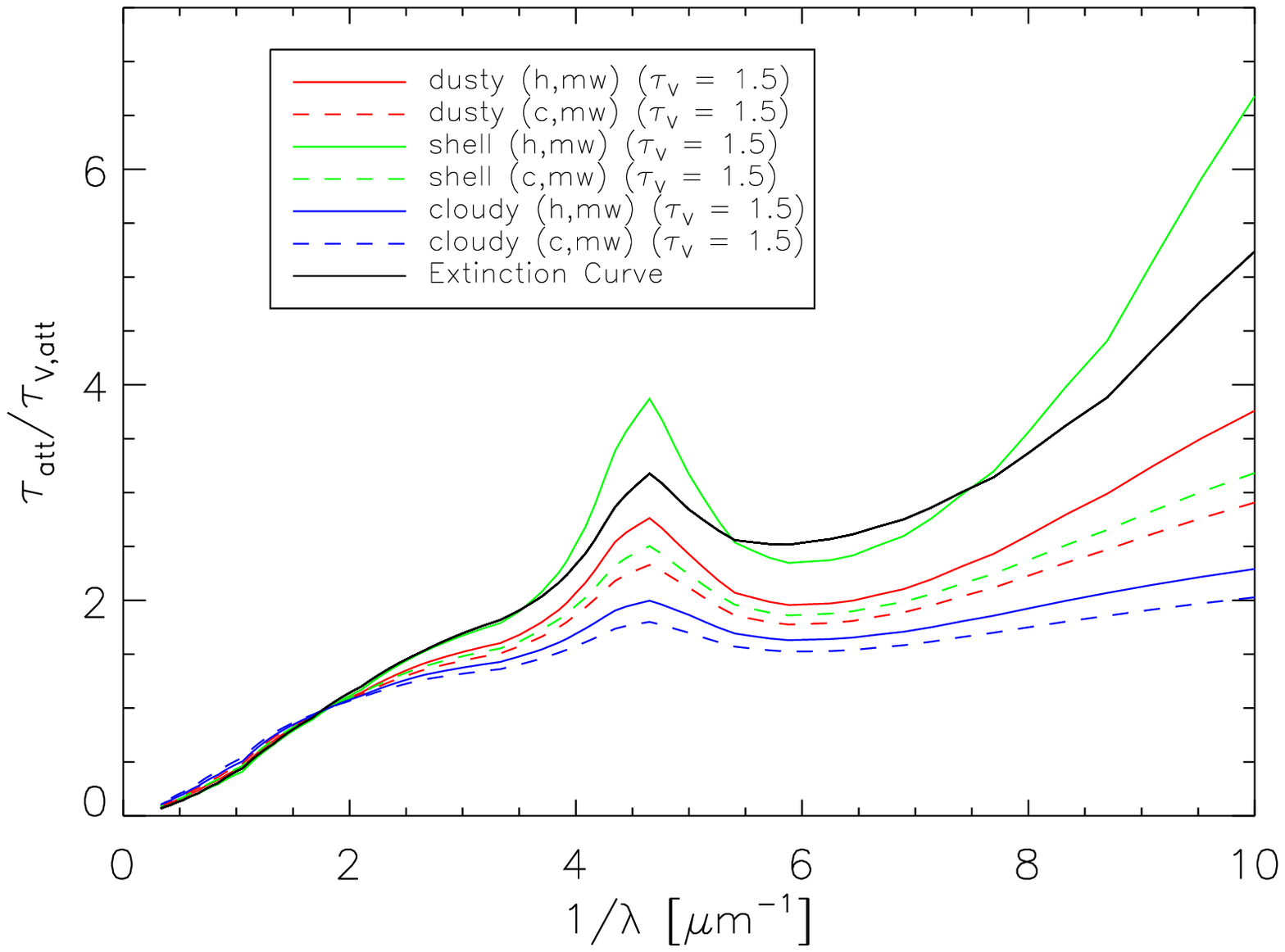} \\
\plottwo{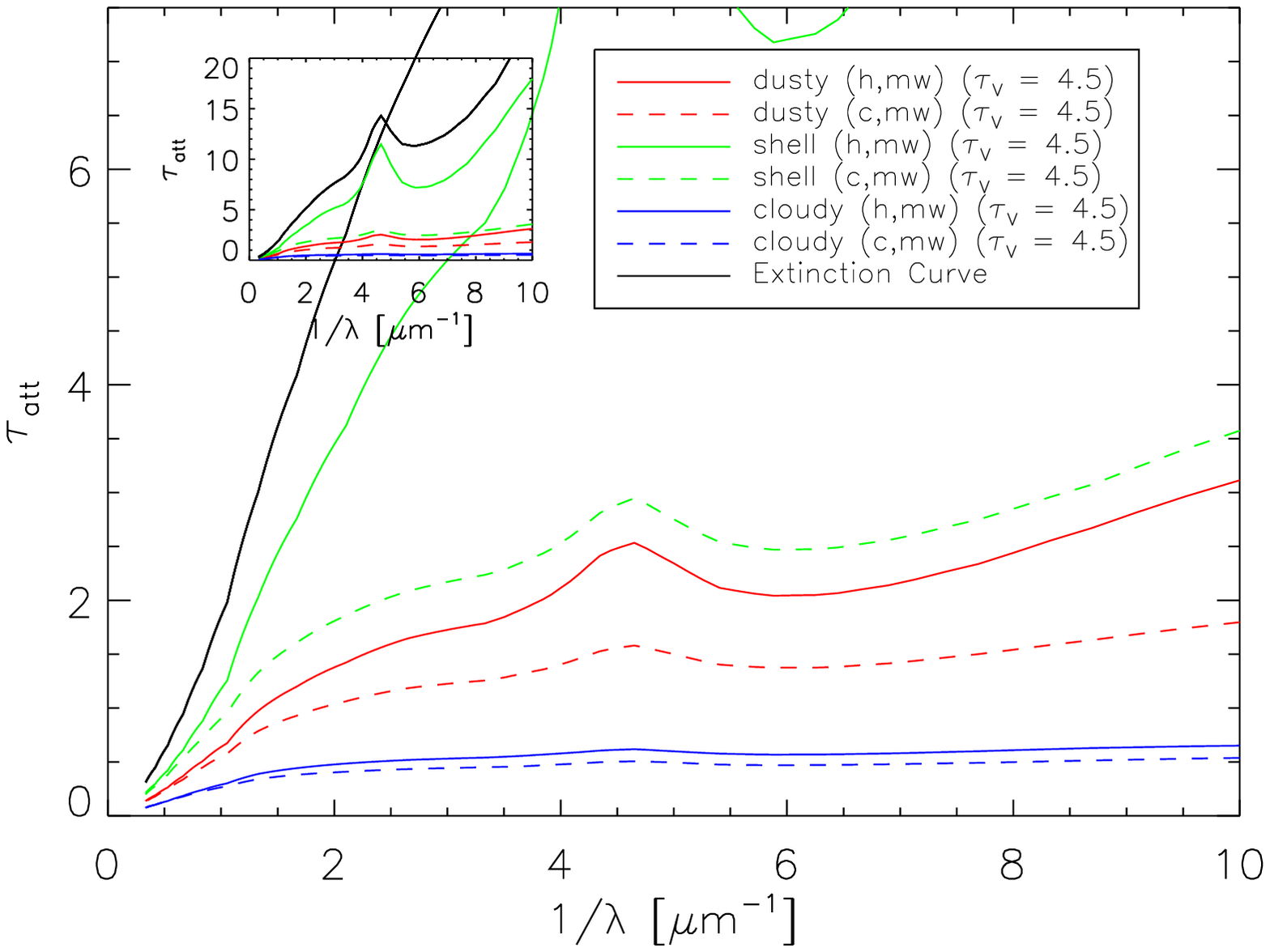}{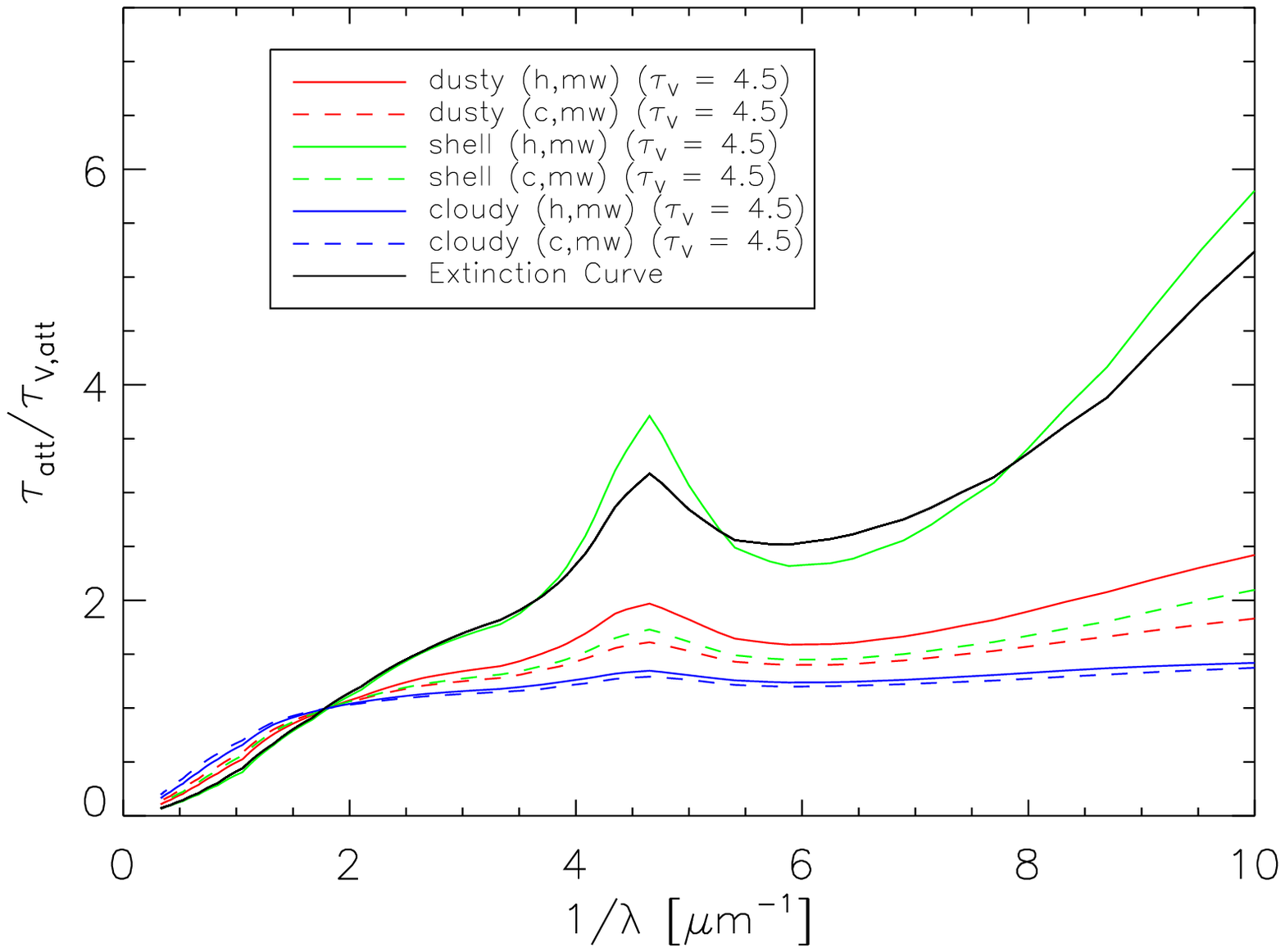}
\caption{The attenuation curves for the Milky Way model runs are
plotted unnormalized (left) and normalized
(right). \label{fig_ext_mw}}
\end{figure*}

\begin{figure*}[tbp]
\plottwo{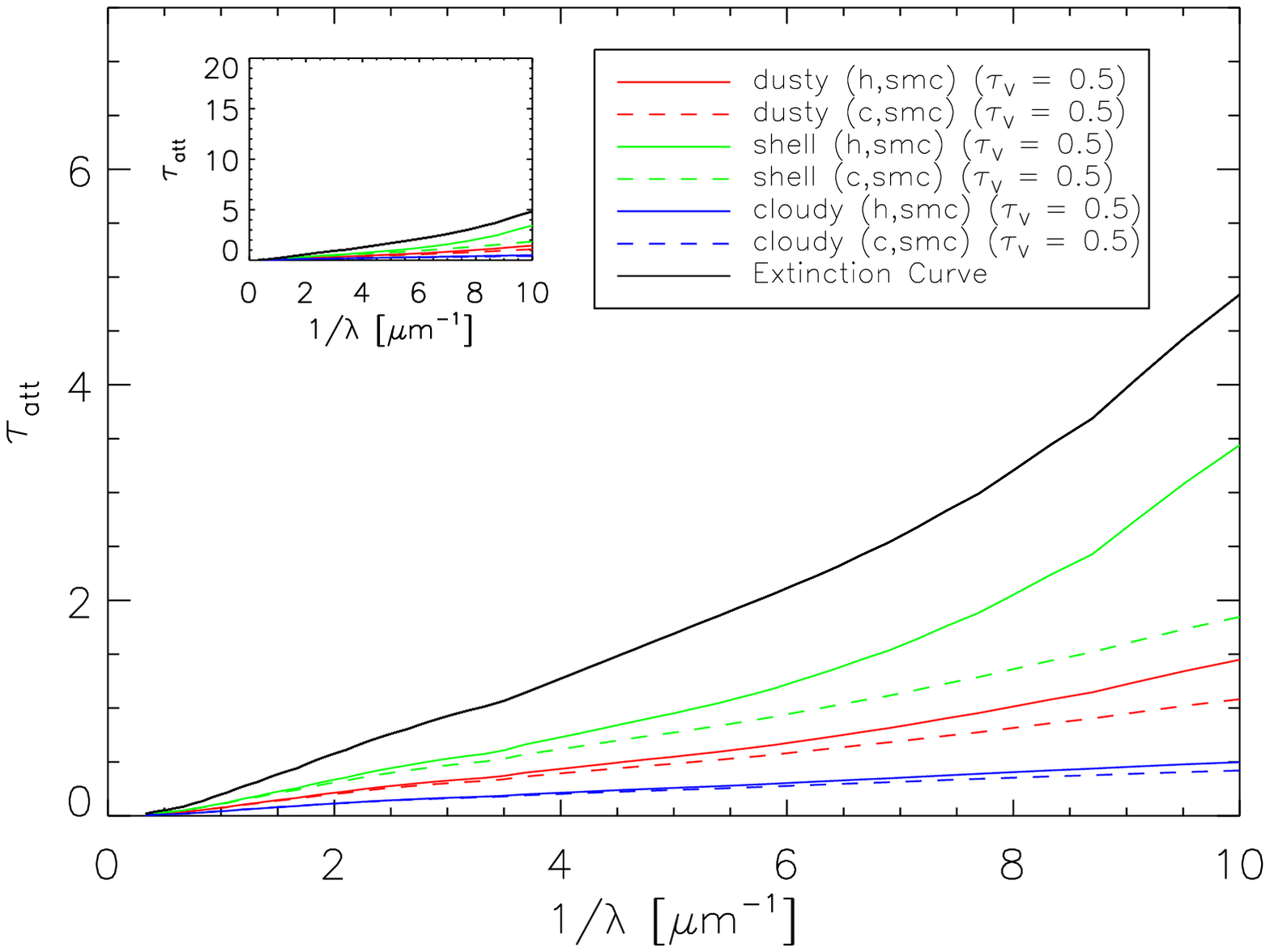}{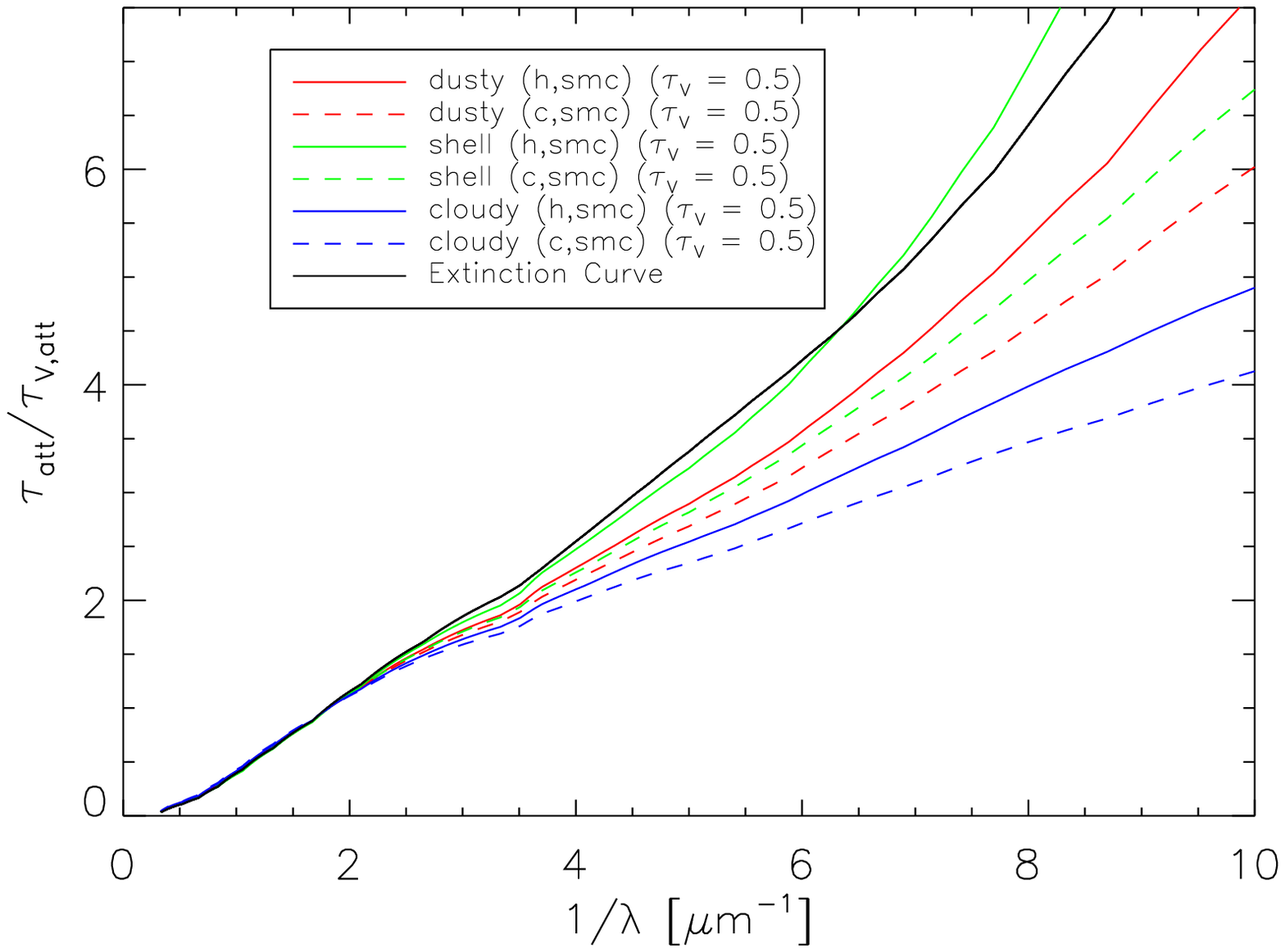} \\
\plottwo{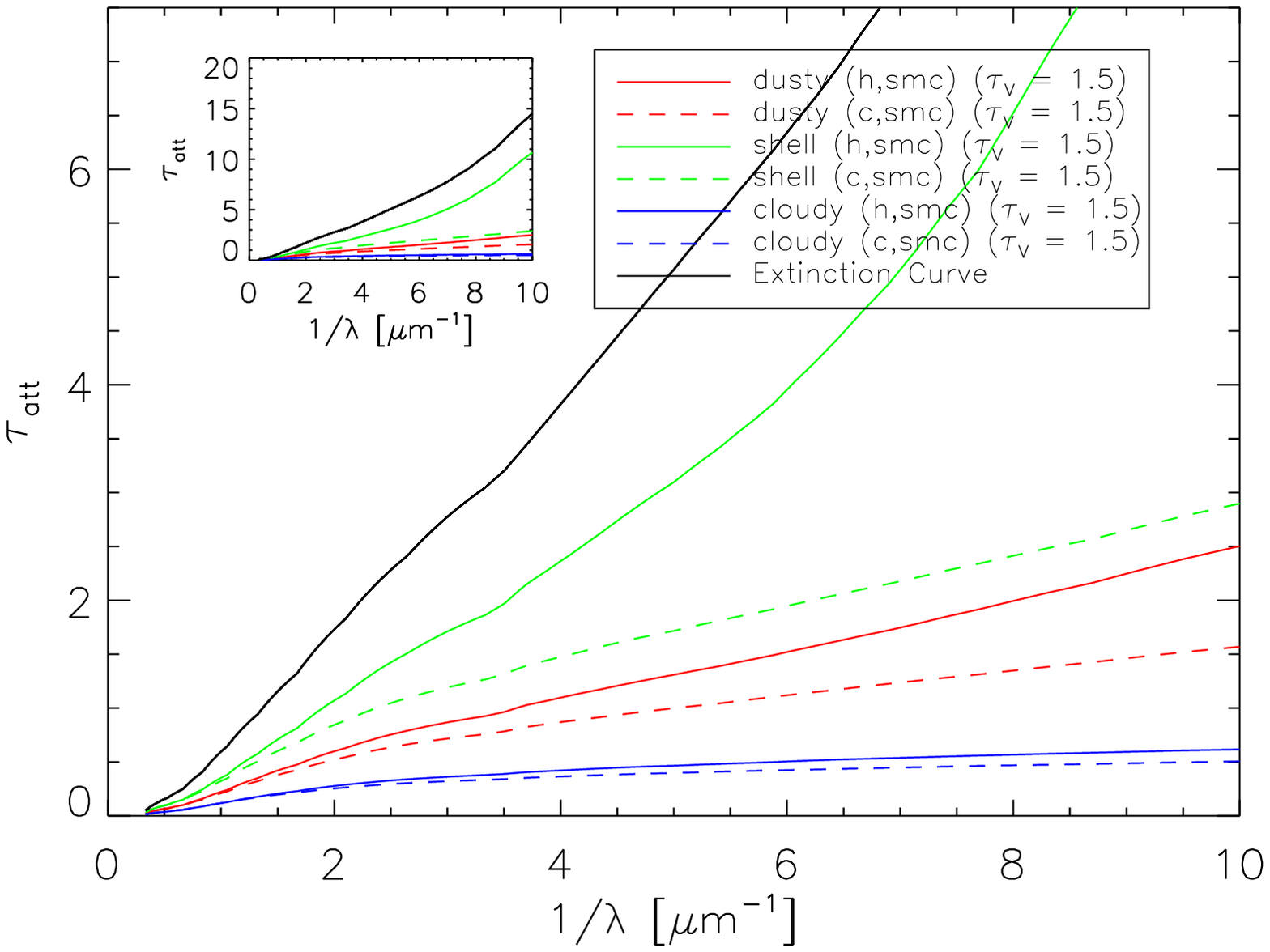}{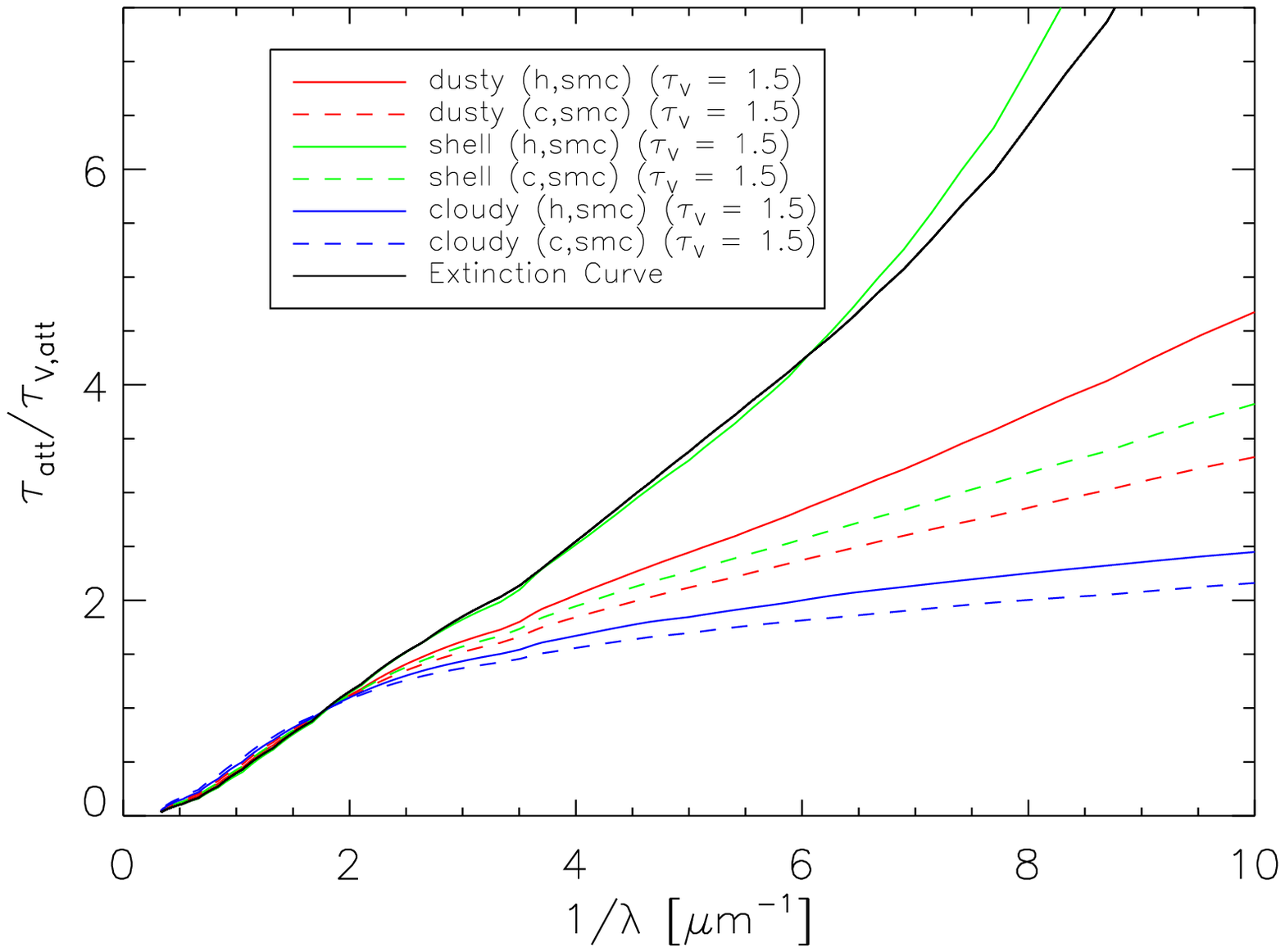} \\
\plottwo{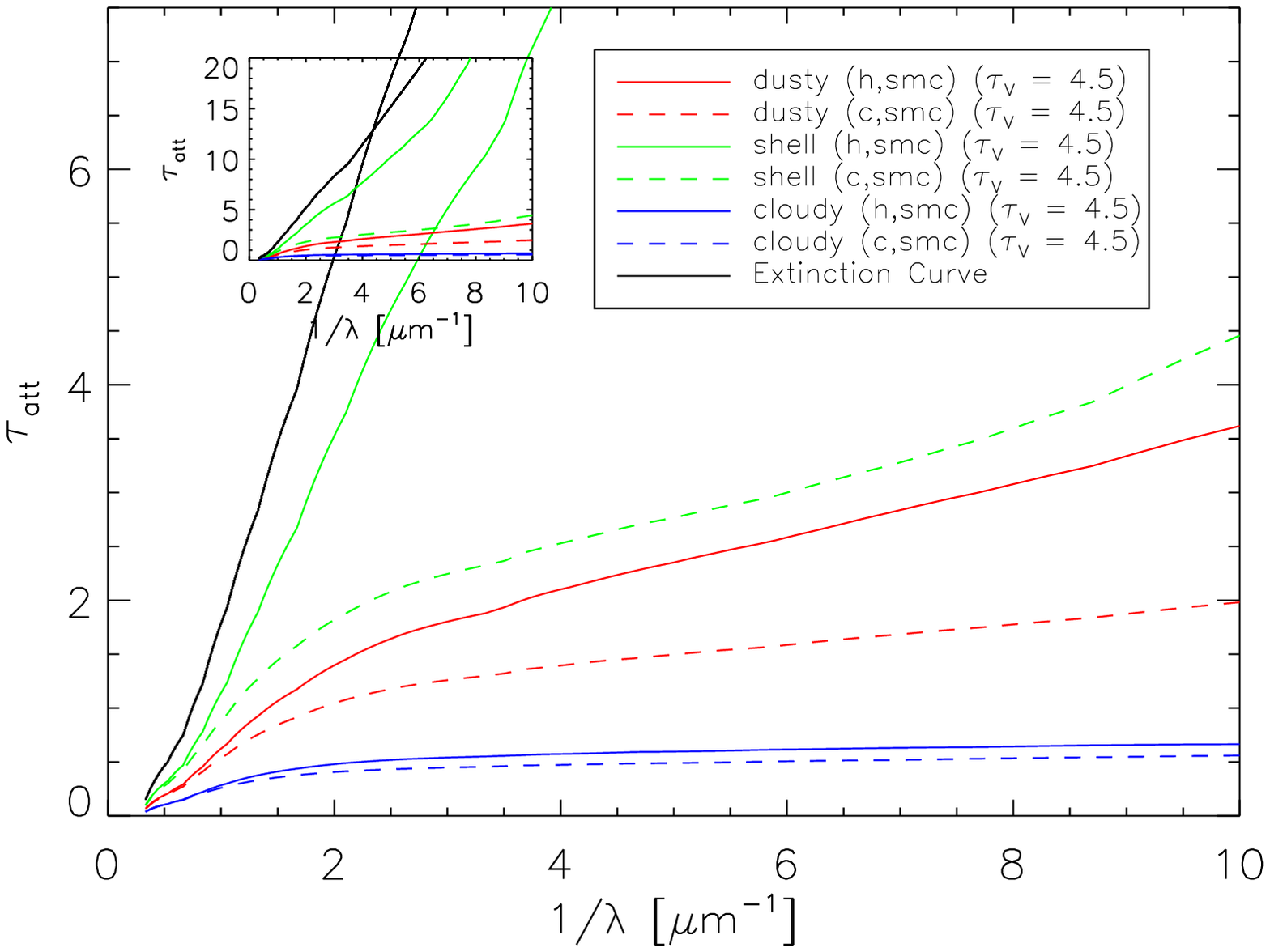}{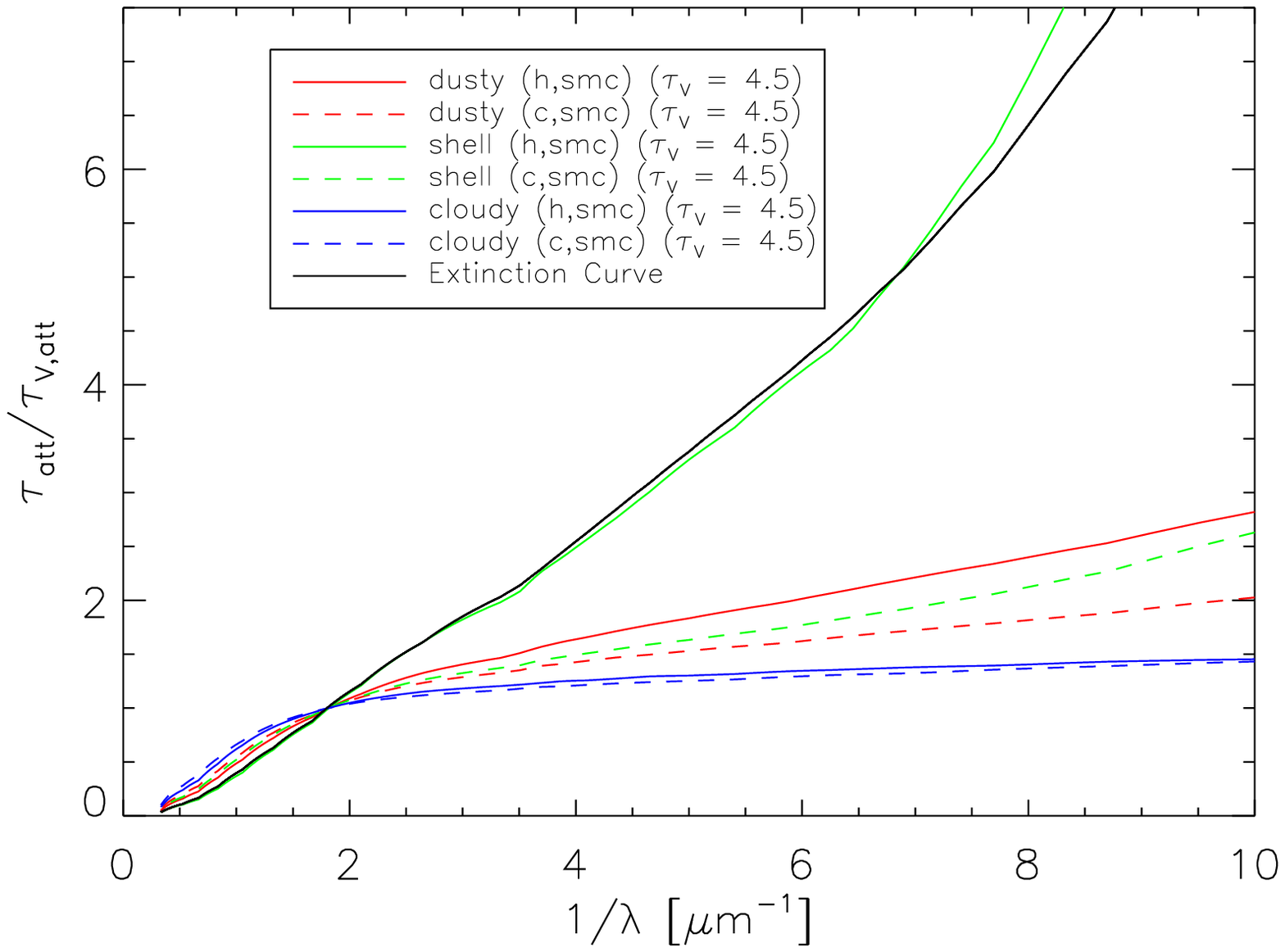}
\caption{The attenuation curves for the Small Magellanic Cloud model
runs are plotted unnormalized (left) and normalized
(right). \label{fig_ext_smc}}
\end{figure*}

In Fig.~\ref{fig_ext_mw}, we present the predicted attenuation optical
depths (see equation 1) as a function of inverse wavelength for MW
dust for three cases, the optically thin case for $\tau_V = 0.5$
(top), the moderately optically thick case $\tau_V = 1.5$ (middle),
and the very optically thick case $\tau_V = 4.5$ (bottom). Also shown
in each panel is the average galactic extinction for the same dust
column densities.  While the left-hand panels show the absolute values
of the respective optical depths, the corresponding right-hand panels
present the same data normalized at V, in order to facilitate the
comparisons of the changes in the wavelength dependence of each
function. The DUSTY, SHELL, and CLOUDY models are contrasted in both
their homogeneous and their clumpy forms. Fig.~\ref{fig_ext_smc}
presents the analogous set of data for SMC dust. Insert boxes are
included to show the behavior of those curves which go out of bounds
vertically on the main graphs.

Several significant conclusions can be drawn from the illustrated
cases.  In all instances, attenuation optical depths for the three
geometries are less than the interstellar extinction produced by the
same column density of dust. This difference is due to the return of
scattered radiation in all cases, exclusively so in the case of the
SHELL geometry, and also due to lower attenuation in cases of stars
only lightly embedded in the dust distribution, which is happening in
the CLOUDY and DUSTY geometries. The descending order from SHELL, to
DUSTY, to CLOUDY geometry can thus be understood. Similarly, clumpy
structures containing the same amounts of dust as corresponding
homogeneous structures always provide less attenuation than the
homogeneous ones at all wavelengths. More importantly, and this
difference is particularly prominent in the SHELL geometry, the clumpy
structures provide a substantially grayer wavelength dependence for
the attenuation than would have been expected from adopting the
original extinction curve. The trend toward gray attenuation increases
strongly with increasing column density, because an increasing
fraction of the total attenuation is caused by clumps which are
optically thick shortward of the V-band. For this same reason, the
relative strength of the 2175~\AA\ UV extinction feature, prominent in
the MW dust, is greatly reduced in clumpy structures. This is in
contrast to the case of the homogeneous SHELL model, which exhibits a
2175~\AA\ feature even stronger than that in the interstellar
extinction (Fig.~\ref{fig_ext_mw}, RHS). This is due to the fact that
the albedo of MW dust displays a minimum coinciding with this feature,
leading to return less scattered light here than at wavelengths
immediately longward or shortward of the feature. Similarly, the
declining dust albedo shortward of 1500~\AA\ causes the predicted
attenuation curve for the homogeneous SHELL model curves to be steeper
in the far UV than the interstellar extinction curve.

It is important to recognize that the slope of the attenuation
functions in the UV for the more realistic embedded and clumpy
geometries is dependent on the total dust column density. The use of a
single attenuation function, for example the widely used ``Calzetti
Attenuation Law'' (\cite{cal97}), for the dereddening and flux
correction of individual starburst galaxies is therefore not justified
in the absence of information about the dust column density. A
comparison of the ``Calzetti Attenuation Law'' with predictions from
our models will be discussed in more detail in Section 4.

\subsection{Scattered Light Component}

\begin{figure*}[tbp]
\plottwo{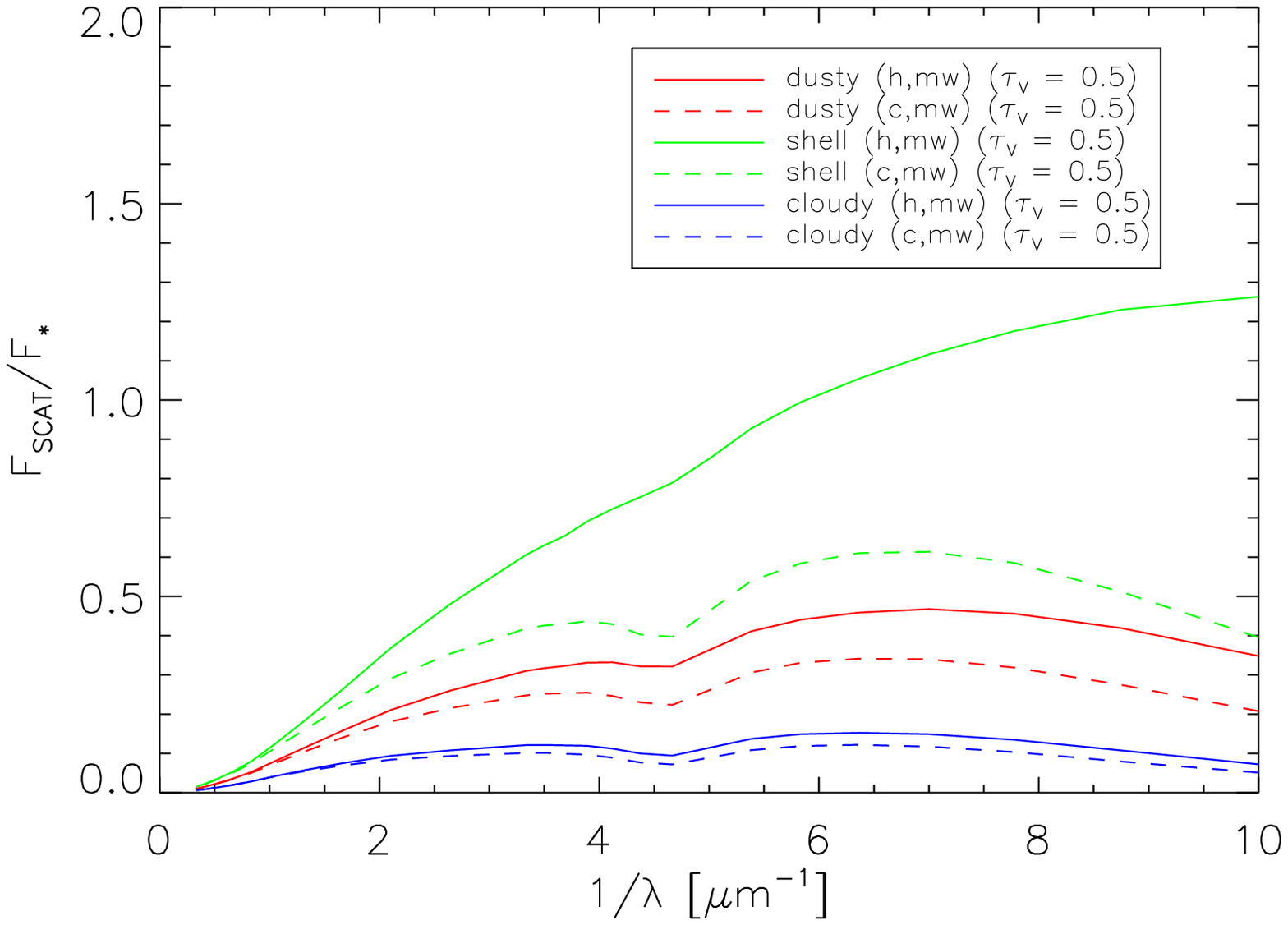}{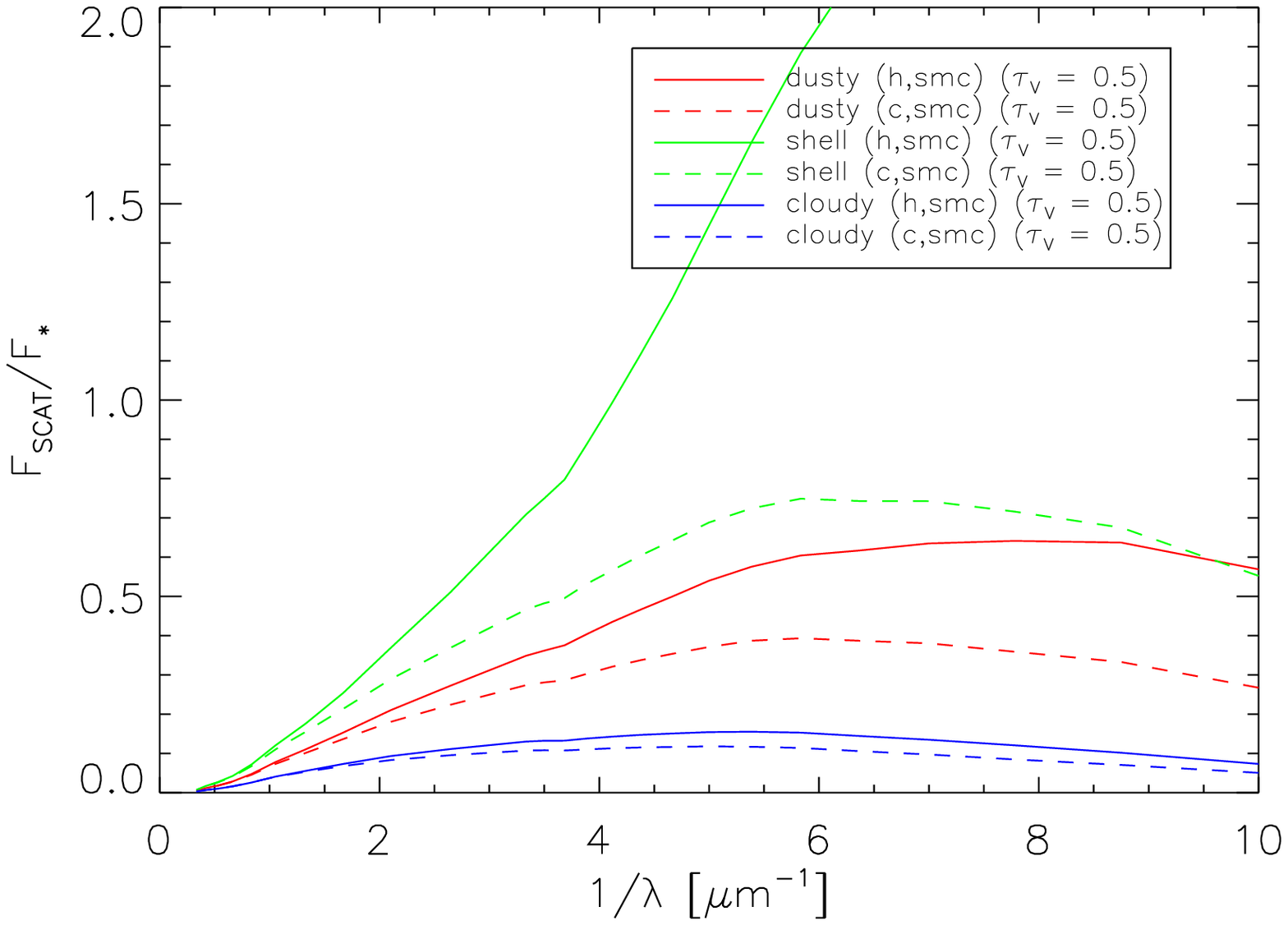} \\
\plottwo{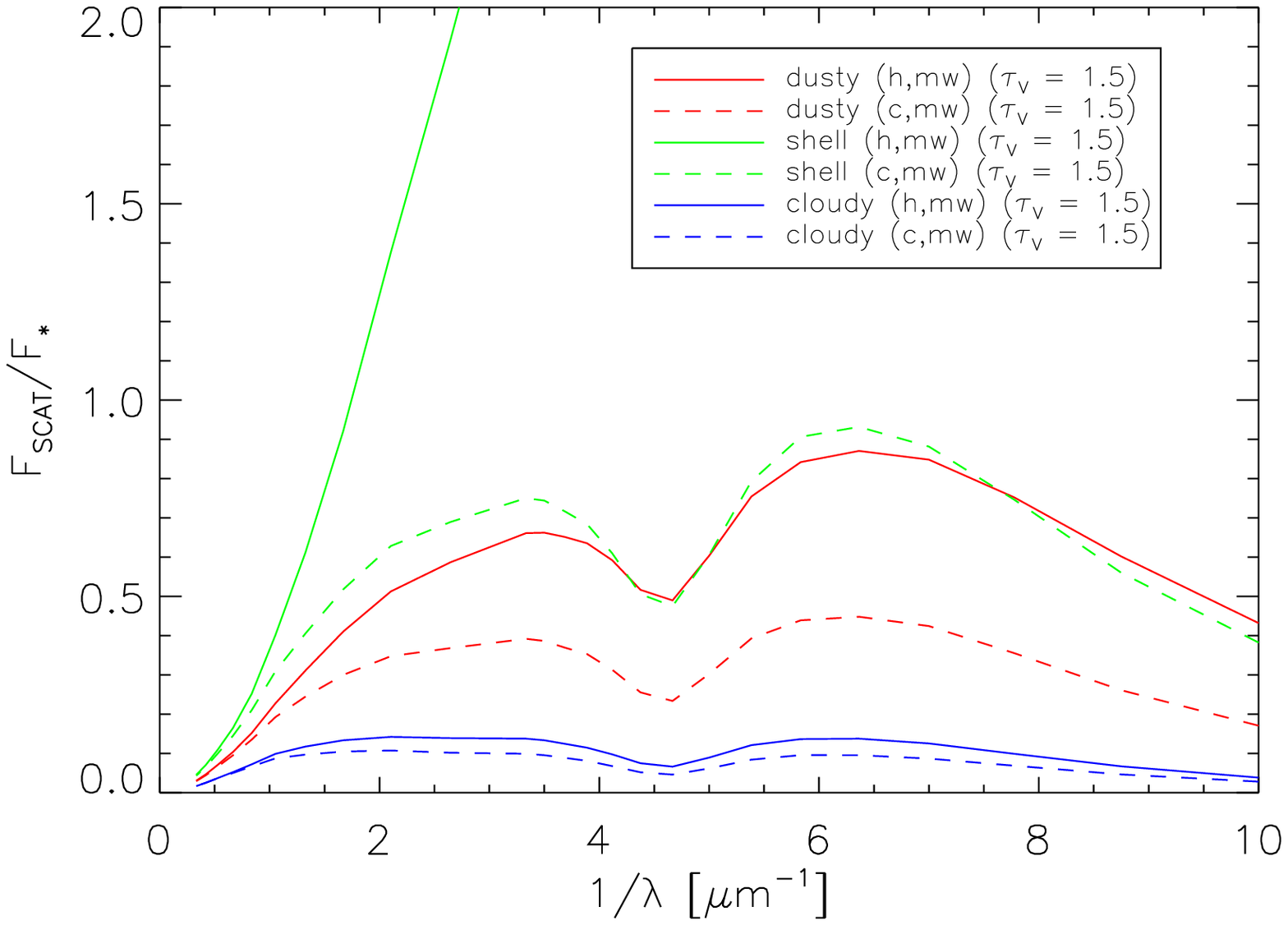}{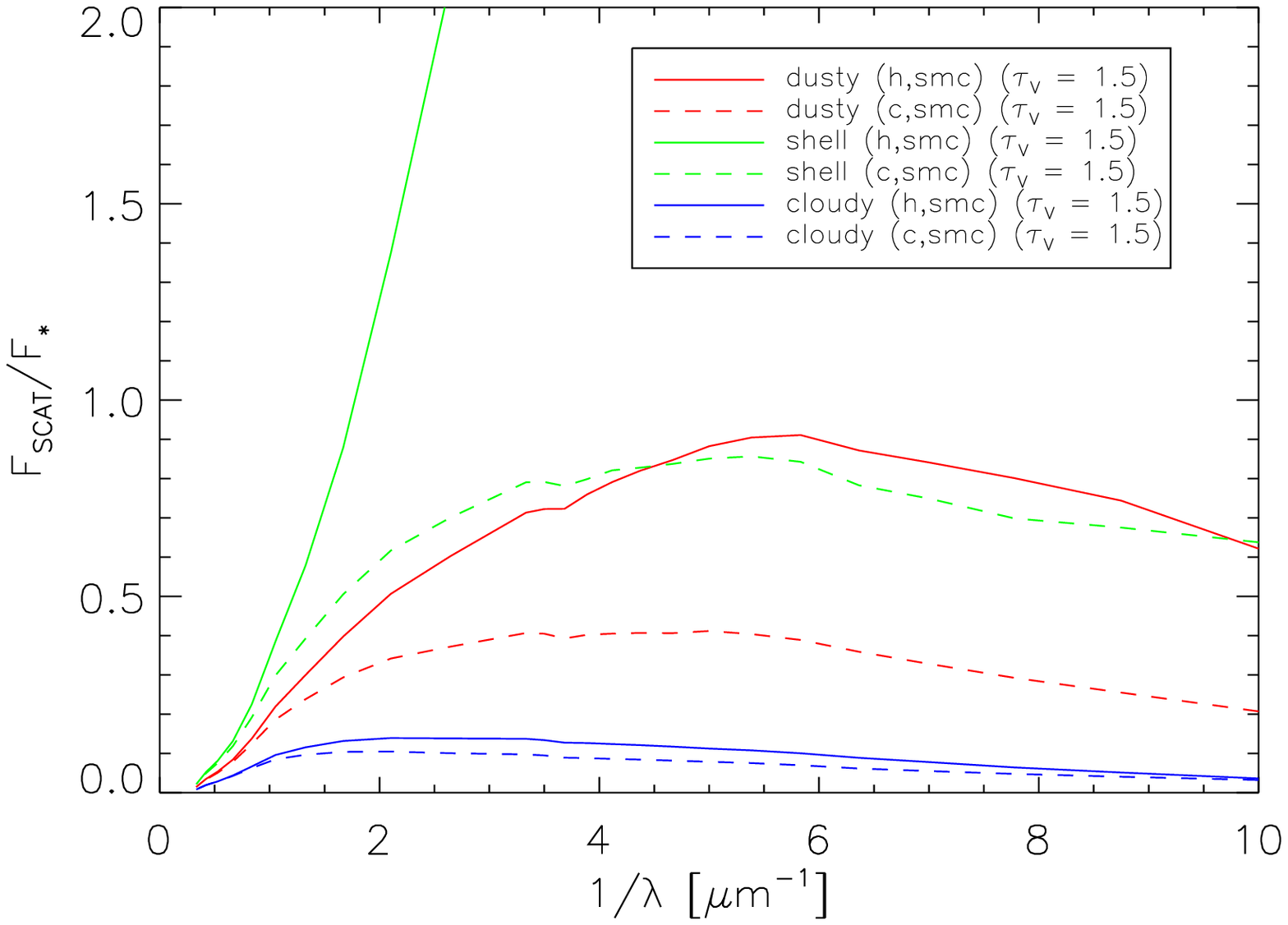} \\
\plottwo{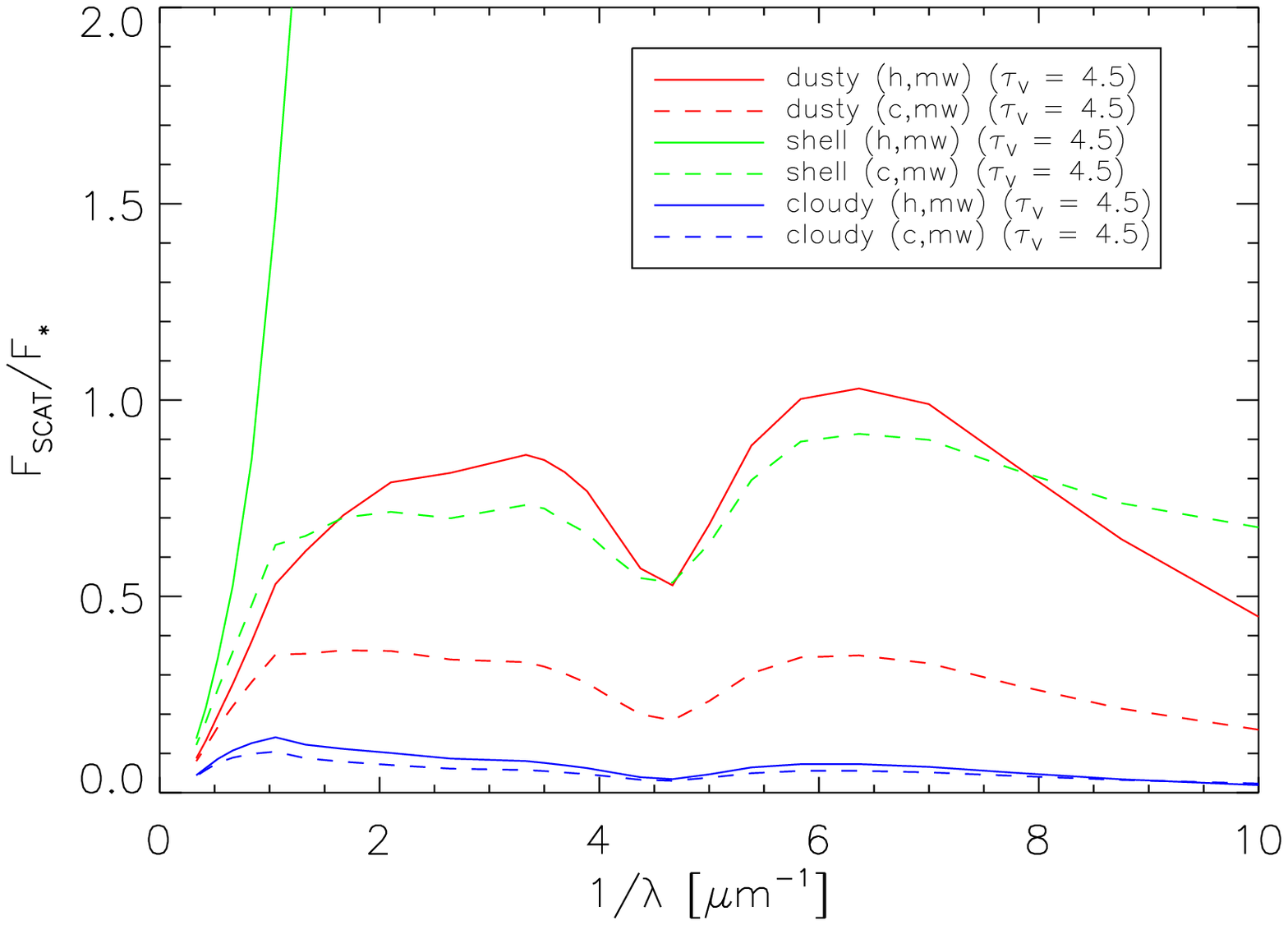}{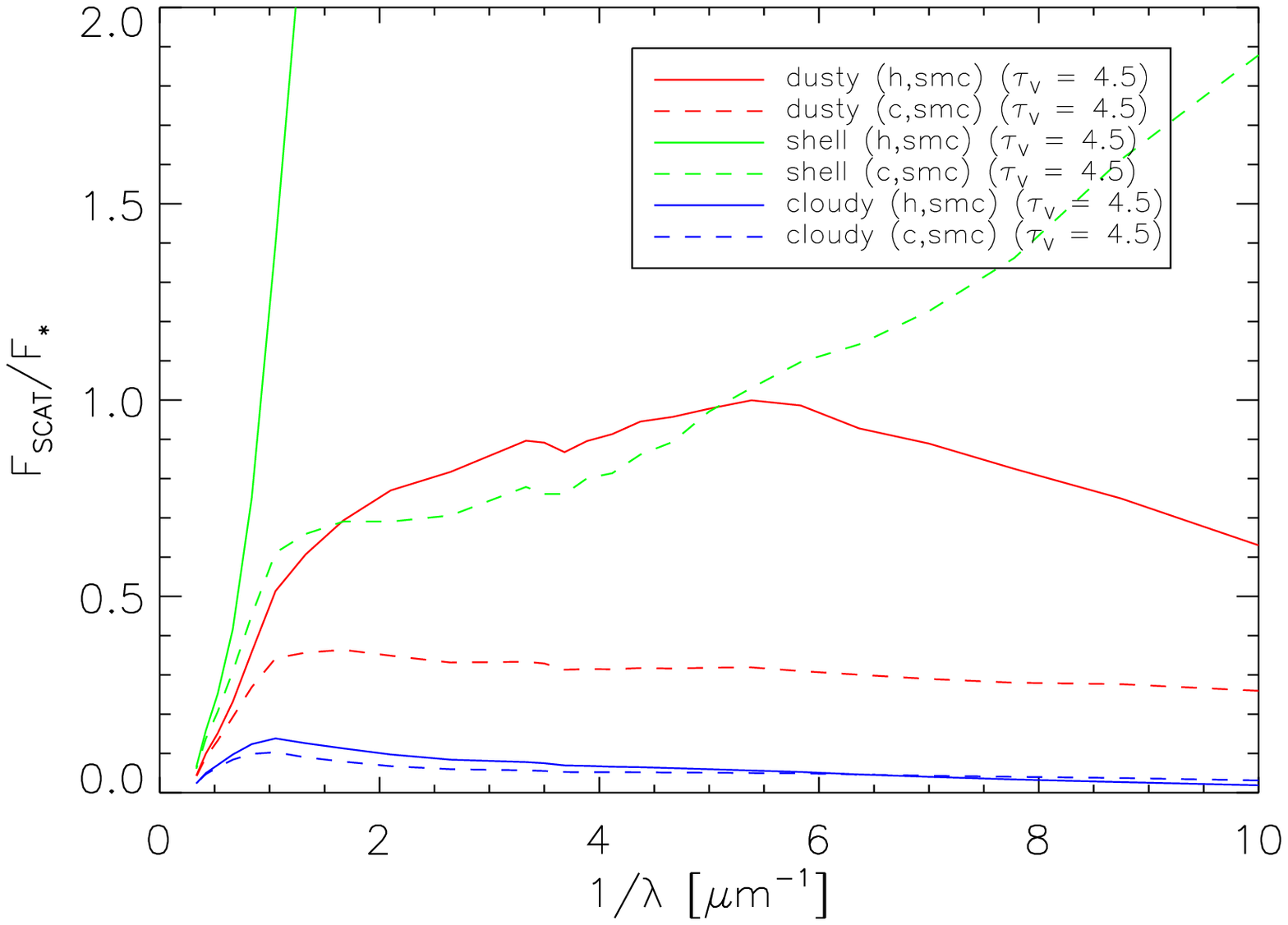}
\caption{The ratio of scattered to stellar flux ($F_{\rm SCAT}/F_*$)
are plotted for the Milky Way (left) and SMC (right) model
runs. \label{fig_scat_flux}}
\end{figure*}

The integrated light of dusty galaxies consists of a certain fraction
of scattered light, with a spectrum similar to that of the integrated
star light. This fraction depends on wavelength, geometry, total dust
column density, and dust scattering properties, as well as viewing
direction in the case of disk galaxies. In the latter, the relative
fraction of scattered light is largest in directions perpendicular to
the disk, this being the direction in which the escape probability is
highest for both stellar and particularly for scattered
radiation. Given the spherical geometry of our models, we are able to
predict the direction-averaged scattered light components. These are
shown in Fig.~\ref{fig_scat_flux} as fractions of the attenuated
stellar flux for three characteristic visual radial optical depths
(0.5, 1.5, 4.5), and for three geometries with both MW and SMC dust,
both clumpy and homogeneous.

As expected from observations of reflection nebulae with embedded
stars (\cite{wit92a}, 1993), the scattered light fraction for
homogeneous SHELL models easily exceeds unity and rises strongly with
increasing optical depth, more as a result of the increasing
attenuation of the starlight, less as a result of increasing scattered
light flux. We view the embedded DUSTY, the clumpy SHELL models, and
the CLOUDY models to be representative of actual geometries of
galaxies. For these, the scattered light fraction appears to be fairly
constant across the UV/visible spectrum and is likely between a
quarter and one half of the total integrated light. In the case of the
MW dust, the depression near $4.6~\micron^{-1}$ is due to the albedo
minimum associated with the 2175~\AA\ extinction bump, and this
feature is actually seen in the spectrum of the diffuse (i.e.\
scattered) galactic light (\cite{wit73}).  The transition from
homogeneous to clumpy structures is associated with a substantial
reduction of the fractional scattered light flux.  This is due to two
causes: Direct starlight escapes more easily from the system, mainly
through gaps in the clump distribution, while the scattered light flux
is diminished, because clumps are mainly illuminated externally with
the scattered light being directed preferentially into the clump
interior where final absorption is more likely. In clumpy media,
clumps in effect become the scattering centers instead of individual
dust grains. As shown in \cite{wit96}, the effective albedo of clumps
is always lower than the input dust albedo.  Overall, the fractional
amounts of scattered light for MW dust and for SMC dust are about the
same for identical dust column densities.

\subsection{Variation in Clumpiness Parameters}

We have emphasized that the models involving clumpy dust distributions
presented in the paper are based upon one particular set of clumpiness
parameters (${\it ff} = 0.15$, $k_2/k_1 = 0.01$, $N = 30$). This set
was found by \cite{wit96} to be closest to that representing the structured
ISM in the Galaxy. Detailed results on how the radiative transfer is affected
when the clumpiness parameters are varied can be found in \cite{wit96}
as well. Here we summarize the principal effects.

The density ratio $k_2/k_1$ and the filling factor ${\it ff}$ together 
control the contributions to the total opacity made by the interclump
medium and the clumps. Our case of ${\it ff}$ = 0.15 and $k_2/k_1$ = 0.01
is intermediate to the case of $k_2/k_1$ near unity, where the interclump
medium controls the system opacity, with results very similar to the
homogeneous case ($k_2/k_1 = 1.0$) and the case of $k_2/k_1 < 10^{-3}$,
where the opacity is controlled by the blocking effect of the clumps
and is then proportional to the filling factor. As the filling factor 
increases, the appearance of the system will approach that of the
homogeneous case.

The grayness of the attenuation (Section 3.3) increases as $k_2/k_1$
decreases for the same amount of dust. Also, the ratio of scattered light
to direct, partially attenuated stellar light (Fig. 8) decreases as the
ratio $k_2/k_1$ decreases, both as a result of the growing probability
of stellar light escaping without any interaction and as a consequence
of the lower scattering ability of a medium dominated by increasingly
optically thick clumps. Finally, for a given amount of dust, the overall 
opacity of the medium will increase as the number of clumps (N) increases.

\section{Discussion}

In this section we discuss which of our models with respect to
geometry, optical depth, and dust type might best be used for the
analysis of observations of different kinds of galaxies.  In
particular, we concentrate on dust attenuation studies of nearby
starburst galaxies and Lyman break galaxies.

\subsection{The ``Calzetti Attenuation Law''}

\begin{figure*}[tbp]
\plottwo{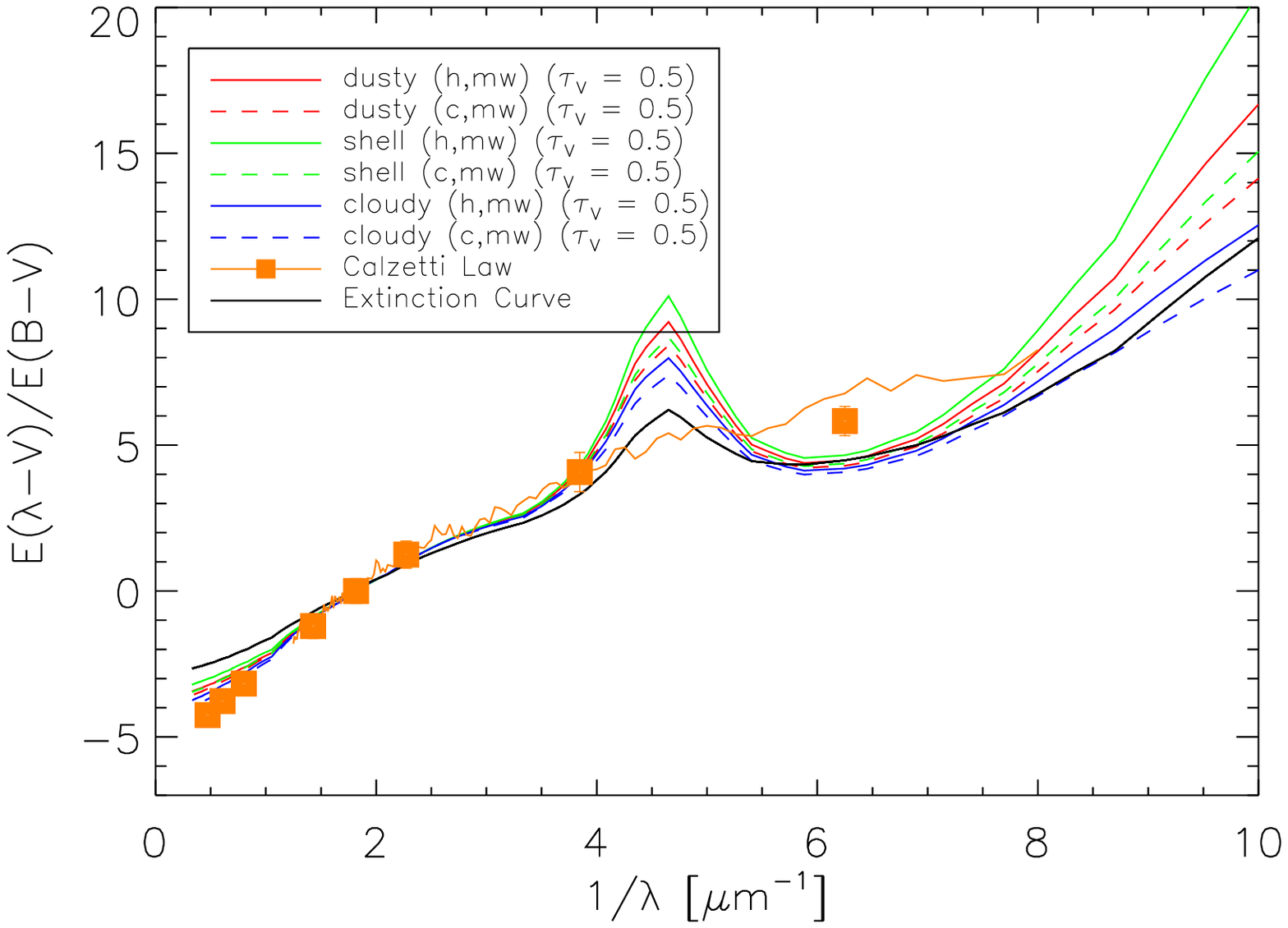}{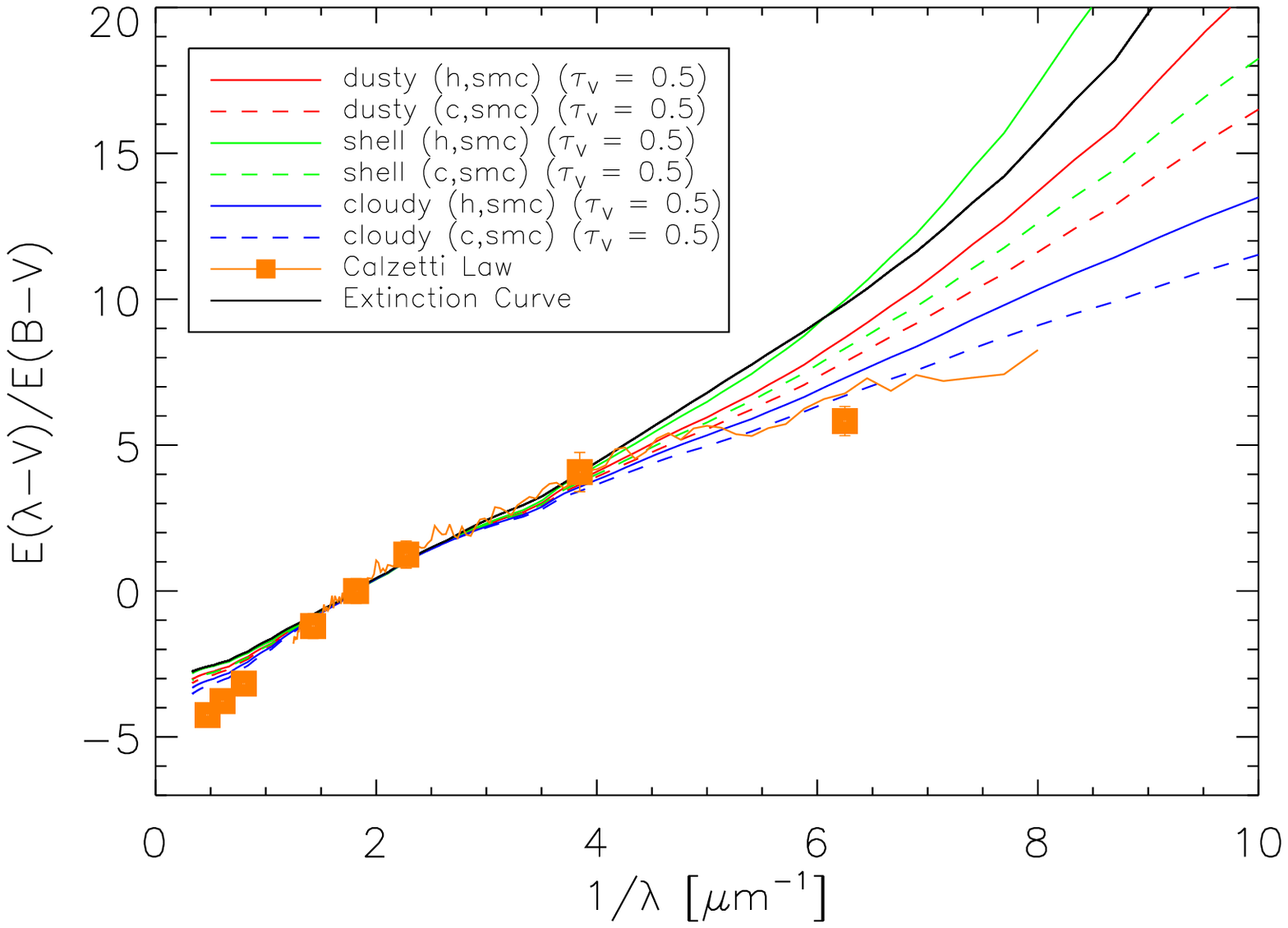} \\
\plottwo{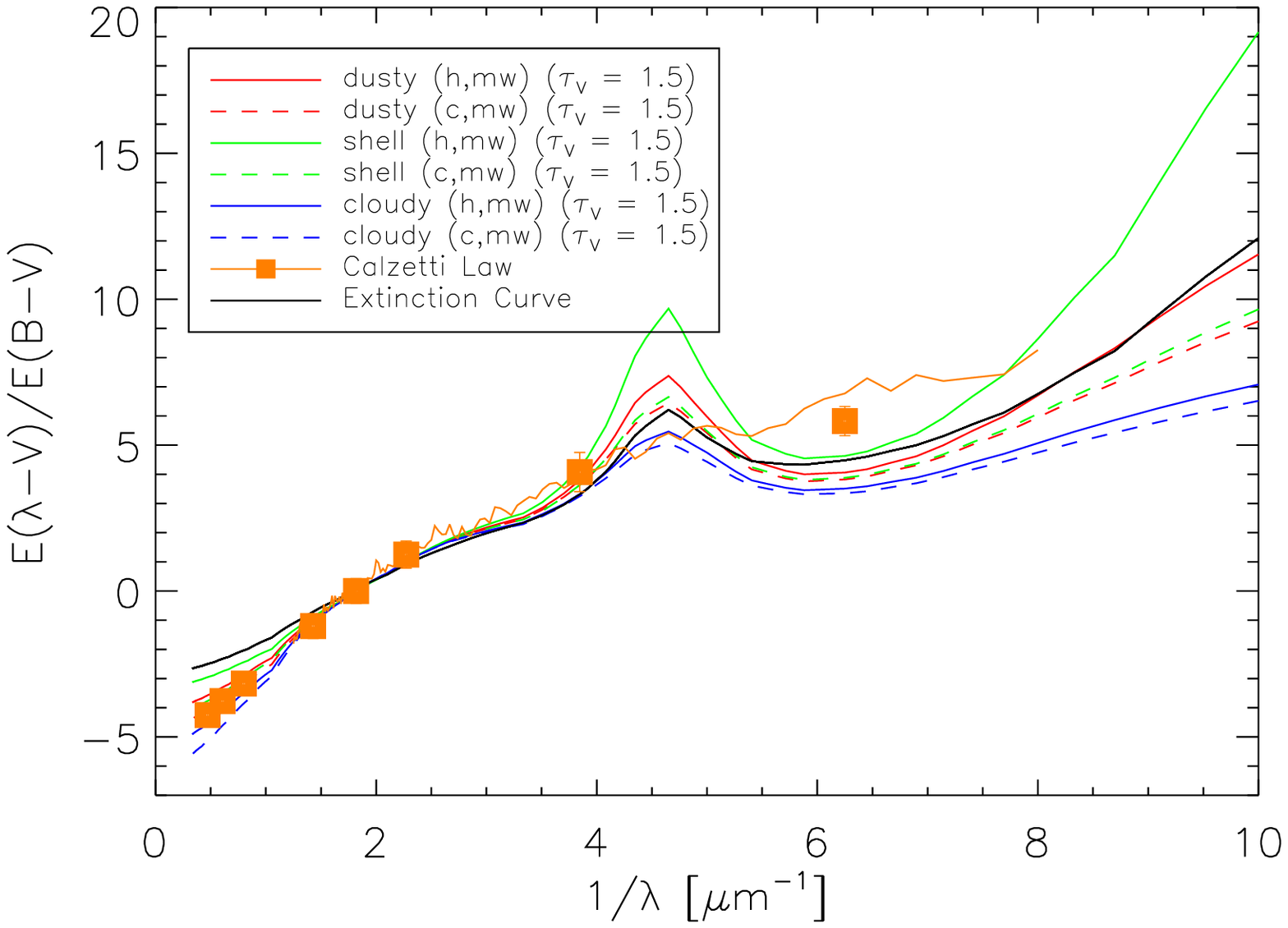}{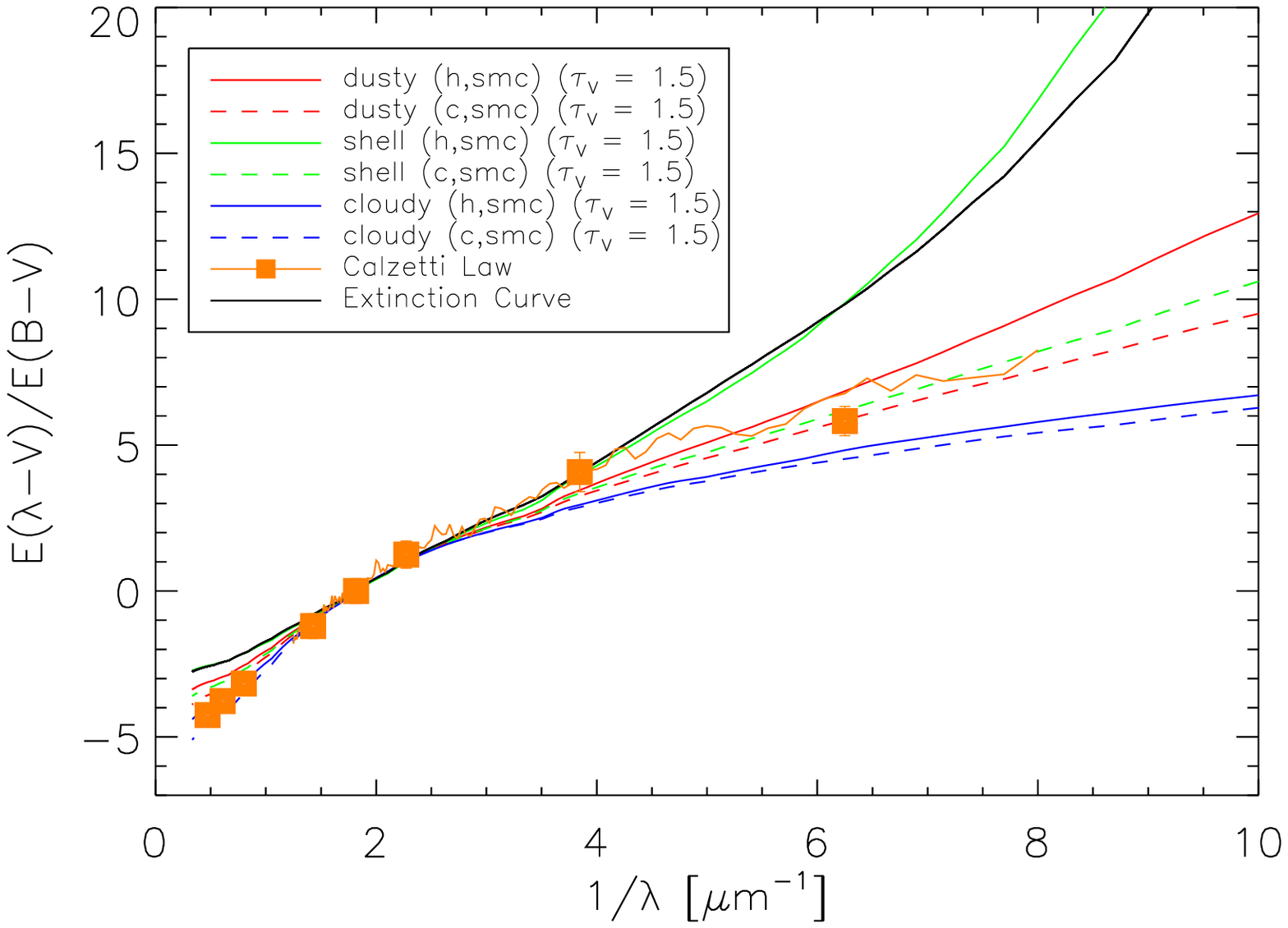} \\
\plottwo{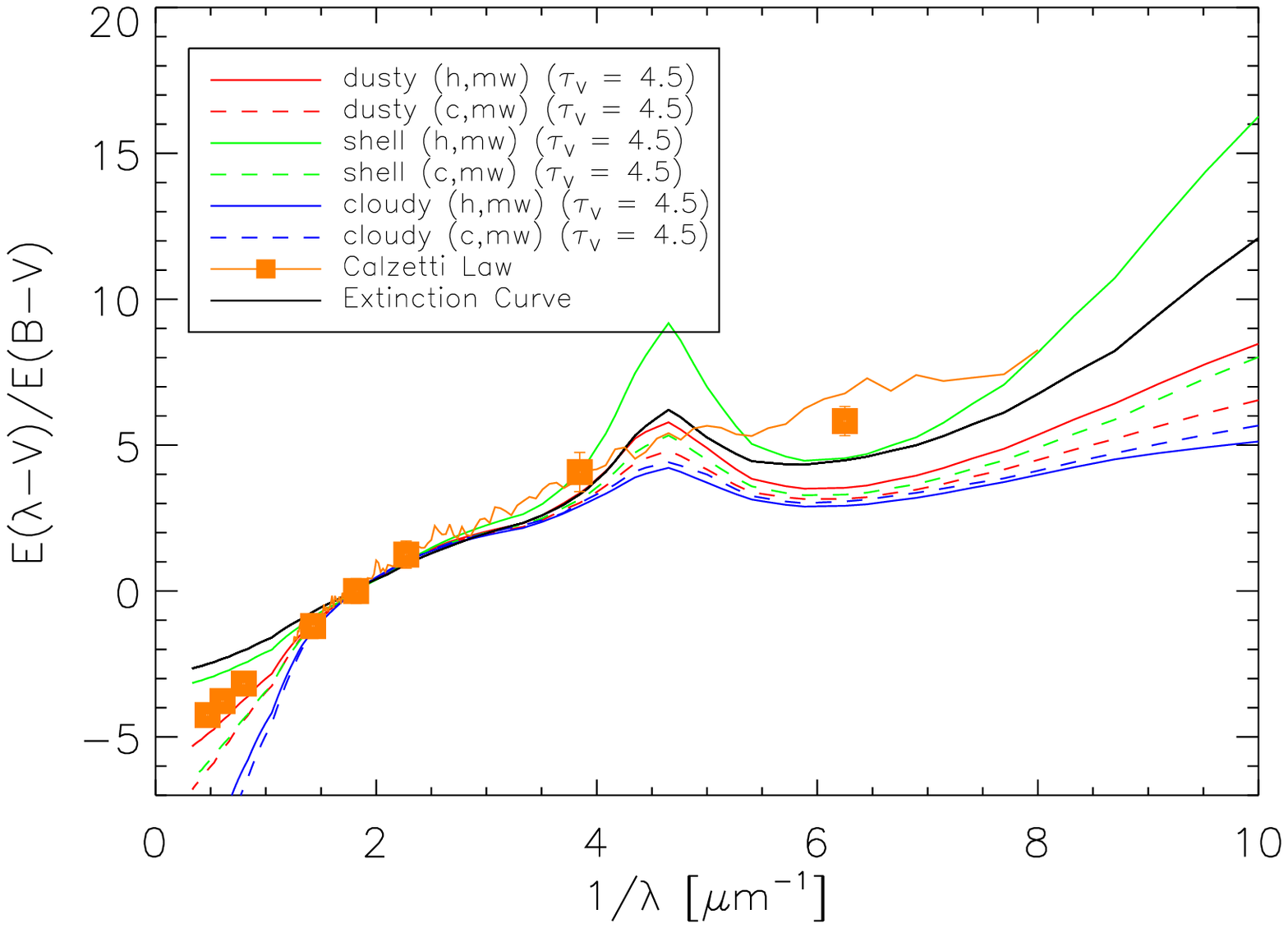}{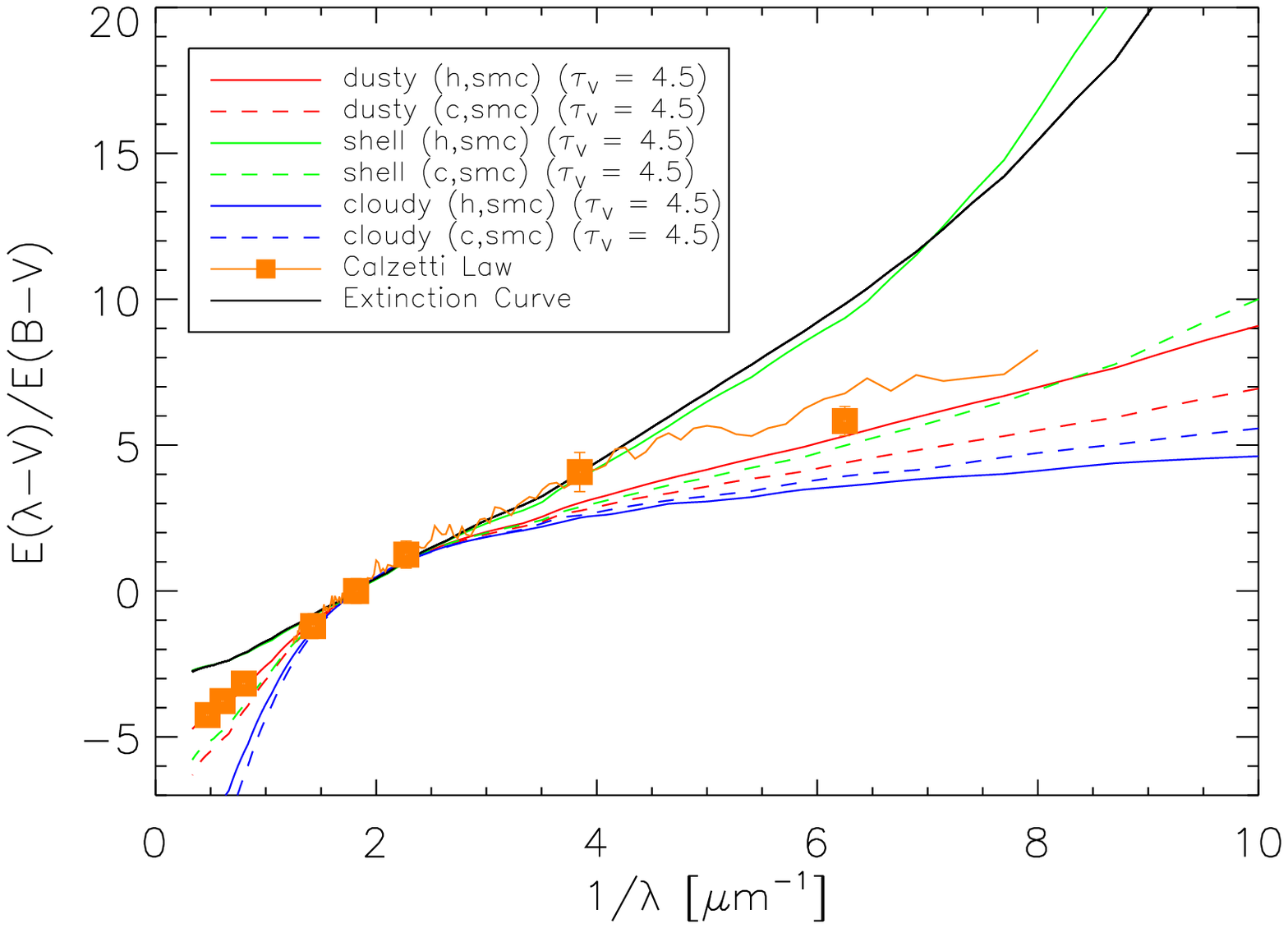}
\caption{The $E(\lambda - V)/E(B-V)$
curves for MW and SMC are plotted along with the Calzetti (1997) data
(filled squares) and the Calzetti et al.\ (1994) data (solid line)
both derived from starburst galaxy observations. \label{fig_ext_cal}}
\end{figure*}

Starburst galaxies exhibit a wide range of dustiness as evidenced by
large variations in the UV spectral index $\beta$, ratios of
recombination line strengths, and ratios of far-IR to UV
luminosities. The UV spectral index $\beta$ was defined by Calzetti et
al.\ (1994) through the relation $F(\lambda) \propto \lambda^{\beta}$
in the range $1250 < \lambda < 2600$ \AA.  By comparing the spectral
energy distributions (SED) of more highly reddened starbursts with
those of barely reddened starbursts, Calzetti et al.\ (1994) derived
the observed wavelength dependence of the dust attenuation in these
systems. The attenuation function takes into account the return of
scattered light into the integrated light of a galaxy and also
reflects radiative transfer effects arising in the three-dimensional
distributions of stars, gas, and dust in these galaxies. This observed
attenuation function, further refined in Calzetti (1997), is referred
to in the literature as the ``Calzetti Attenuation Law'' or ``Calzetti
Law''. This attenuation due to internal dust must be clearly
distinguished from the wavelength dependence of interstellar
extinction, which is the attenuation due to absorption {\em and}
scattering by a screen of dust located between a distant source and an
observer.

The principal characteristic of the ``Calzetti Law'' is the absence of
a 2175~\AA\ feature. Gordon et al.\ (1997) showed that the ``Calzetti
Law'' can be reproduced, if the radiative transfer in starburst
galaxies is occurring in a medium with intrinsic dust properties
closely resembling that of SMC dust.  Also, the large range of
observed reddenings, as measured by variations in the spectral index
$\beta$, appears to require efficient attenuation geometries such as
provided by our SHELL or DUSTY geometries.

In Fig.~\ref{fig_ext_cal} we compare attenuation values derived from
starburst galaxies by Calzetti (1997) with our model attenuation
functions. In order to facilitate this comparison, models and
observations are renormalized to $E(\lambda - V)/E(B-V)$.  In order to
transform the values in Table~5 of Calzetti (1997), we have used the
result of Calzetti (1997) that the stellar continuum suffers 40\% the
reddening that the gas suffers, i.e.\ $E(B-V)_{\rm stellar} =
0.4E(B-V)_{\rm gas}$.  The data point capable of discriminating
between different models most clearly is the one near
$6~\micron^{-1}$, derived from the far-UV slope in the reddened SEDs
of starbursts. No model with MW dust matches this data point. In
addition, the lack of a 2175~\AA\ feature in the observations compared
to the prediction of a 2175~\AA\ feature in the models with MW dust
argues strongly against the presence of MW dust in starburst
galaxies. This will be discussed further in Section~\ref{sec_bump}.

Among SMC dust models, the same point does discriminate between dust
column densities and clearly favors the clumpy SHELL model for $\tau_V
= 1.5$.  As is apparent from Fig.~\ref{fig_ext_smc}, such a model
predicts an attenuation optical depth of about $\tau_{\rm att} = 2.0$
at $6~\micron^{-1}$, and, according to Fig.~\ref{fig_ebv_tauV}, a
color excess $E(B-V) = 0.22$. An attenuation optical depth of 2.0
implies that the flux at $6~\micron^{-1}$ has been reduced by a factor
7.4 on average in this UV-selected sample of starburst galaxies . We
need to re-emphasize that there is no universal attenuation function
and that the attenuation factor of 7.4 corresponds simply to the
average of the sample of UV/optically-selected starbursts studied by
Calzetti et al.\ (1994). Our average attenuation factor of 7.4
compares well with the luminosity-weighted mean absorption factor of
5.4 derived by Meurer et al.\ (1999) for the same sample. We discuss
the UV flux reduction factor and the uncertainties associated with its
determination further in Section~\ref{sec_uv_atten}.

\vspace*{0.05in}
\epsscale{0.45}
\begin{center}
\plotone{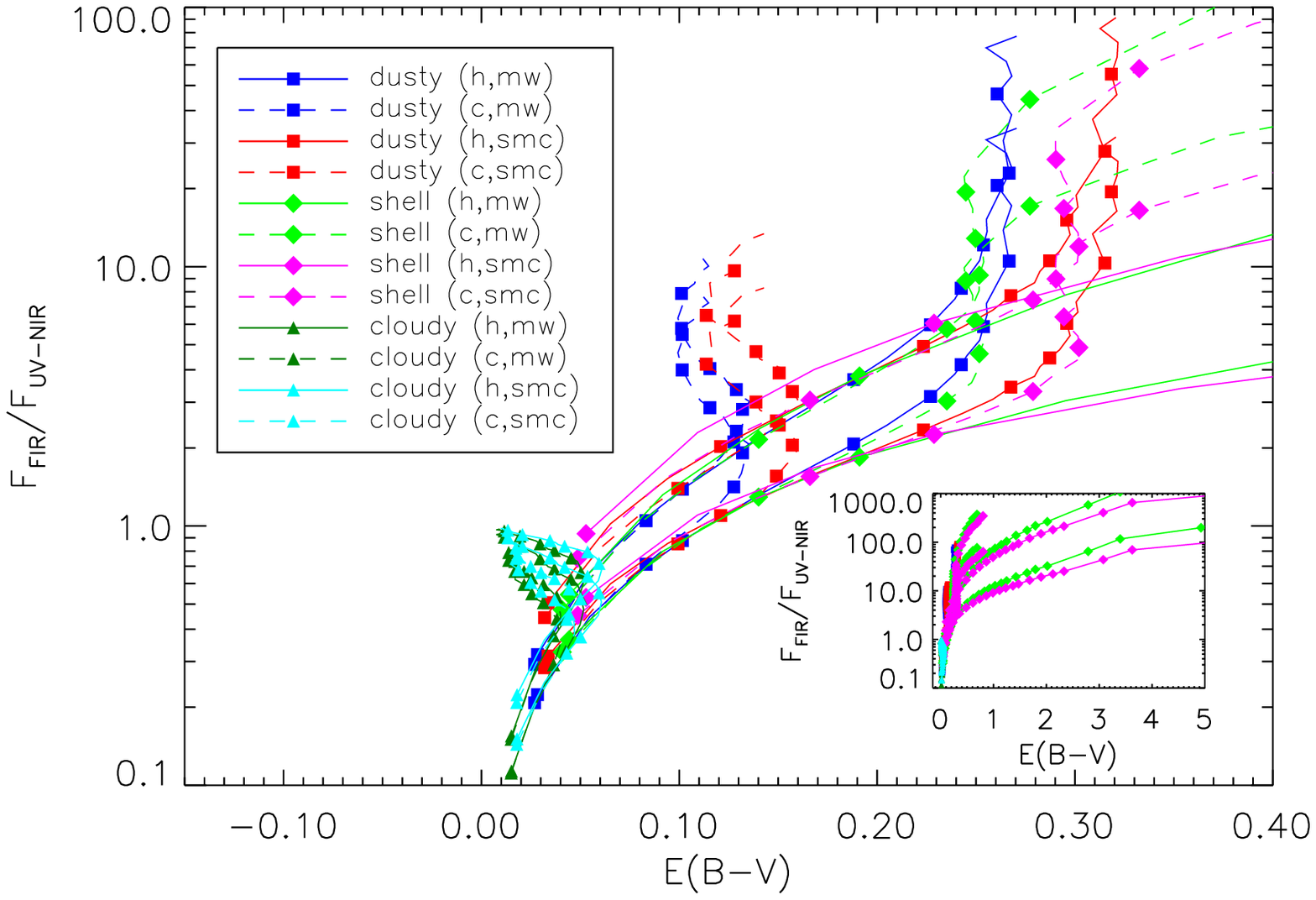}
\end{center}
\figcaption{The ratio of FIR to UV-NIR flux ($F_{\rm FIR}/F_{\rm
UV-NIR}$) is plotted versus $E(B-V)$.  Each model curve is shown
twice, once assuming a 1 million year old constant star formation SED
and once assuming a 1 billion year old constant star formation SED.
The symbols are plotted every third model $\tau_V$ point.
\label{fig_ratio}}
\vspace*{0.05in}

\subsection{The Far-IR to Optical Flux Ratio}

It came as an initial surprise during the IRAS mission that far-IR
ultraluminous and by inference, very dust-rich, galaxies in most
instances exhibited optical colors which were barely distinguishable
from those of normal galaxies of similar population
type. \cite{wit92b} showed that this was due to the fact that, given
embedded dust geometries, the optical colors were determined by the
least attenuated stars, while the far-IR flux was due to the
absorption of light from stars which contribute little or nothing to
the optical flux. The present calculations were extended to dust
column densities corresponding to $\tau_V = 50$ instead of 15 in the
case of \cite{wit92b}. In Fig.~\ref{fig_ratio}, we show the resulting
ratios of the far-IR flux, integrated over all frequencies at which
dust emission occurs, to the integrated UV/optical/near-IR stellar
flux. We reproduce the most extreme observed ratios near 100 with the
clumpy SHELL model of $\tau_V = 50$ with a maximum reddening near 0.8
in $E(B-V)$ and only slightly lower ratios with the homogeneous DUSTY
models with a reddening less than 0.3 in $E(B-V)$. Only the
homogeneous SHELL, as expected for a simple screen, produces very
large amounts of reddening of $E(B-V) \sim 4.0$ and $F_{\rm
IR}/F_{UV-NIR} \sim 1000$ for $\tau_V = 50$. The fact that most IR
luminous galaxies appear to be relatively blue seems to indicate that
clumpy dust structures prevail in most of them. This is also confirmed
by the correlation between the ratio $F_{\rm IR}/F_{\rm F160BW}$ and
the spectral index $\Delta\beta$ for different geometries and dust
types, shown in Fig.~\ref{fig_UVratio}.

\subsection{The UV Flux Attenuation Factor \label{sec_uv_atten}}

A major aim of studies of high-z starbursts (or Lyman-break galaxies)
is to determine the epoch of peak star formation in the history of the
Universe, if such a peak indeed exists. The outcome of such an
investigation depends critically upon the corrections applied for dust
attenuation in the rest-frame UV continuum spectrum of the observed
galaxies, given that these systems contain substantial amounts of
internal dust ( \cite{dic97}; \cite{meu97};
\cite{hec98}; \cite{meu99}; \cite{ste99}).  In
addition, the upcoming NASA GALEX mission is to produce about 100,000
UV galaxy spectra in order to probe the global history of star
formation, and the observed UV spectral slope is be be related
uniquely to the attenuation factor by which the UV flux has been
reduced by dust internal to the galaxies.

Our Figures~\ref{fig_ext_mw} and \ref{fig_ext_smc} already indicated
some of the problems to be expected in such an approach. The
attenuation functions are geometry dependent and they become
increasingly gray as the dust column density increases.  This means
that a heavily attenuated starburst may exhibit a steep UV slope in
its SED, resembling that of a barely reddened galaxy, but show a
greatly reduced UV flux, corresponding to a system with little star
forming activity.  The chances for underestimating the star formation
rate in early epochs of the Universe are therefore severe, if dust is
abundant generally in the early galaxies.

\begin{figure*}[tbp]
\plottwo{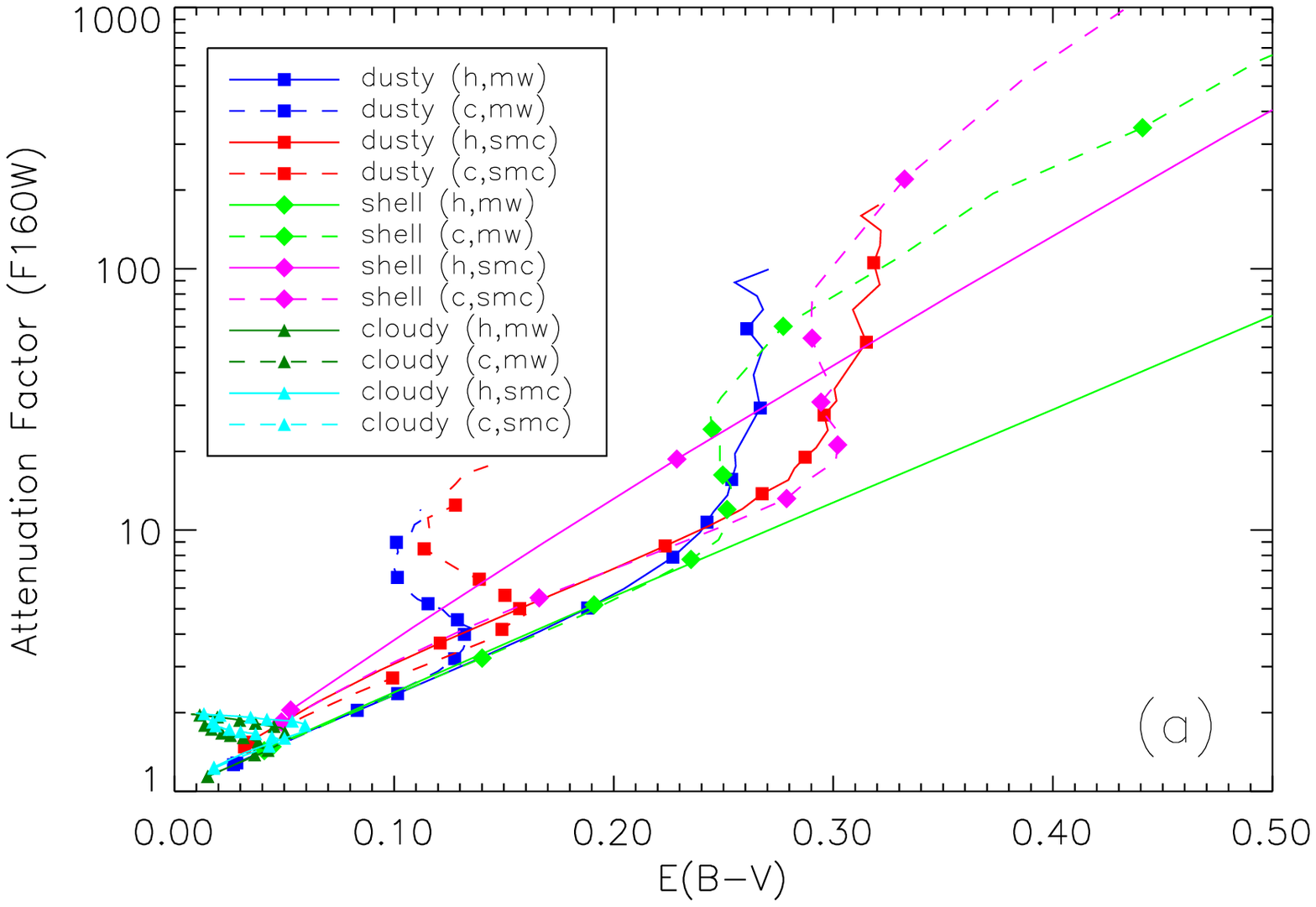}{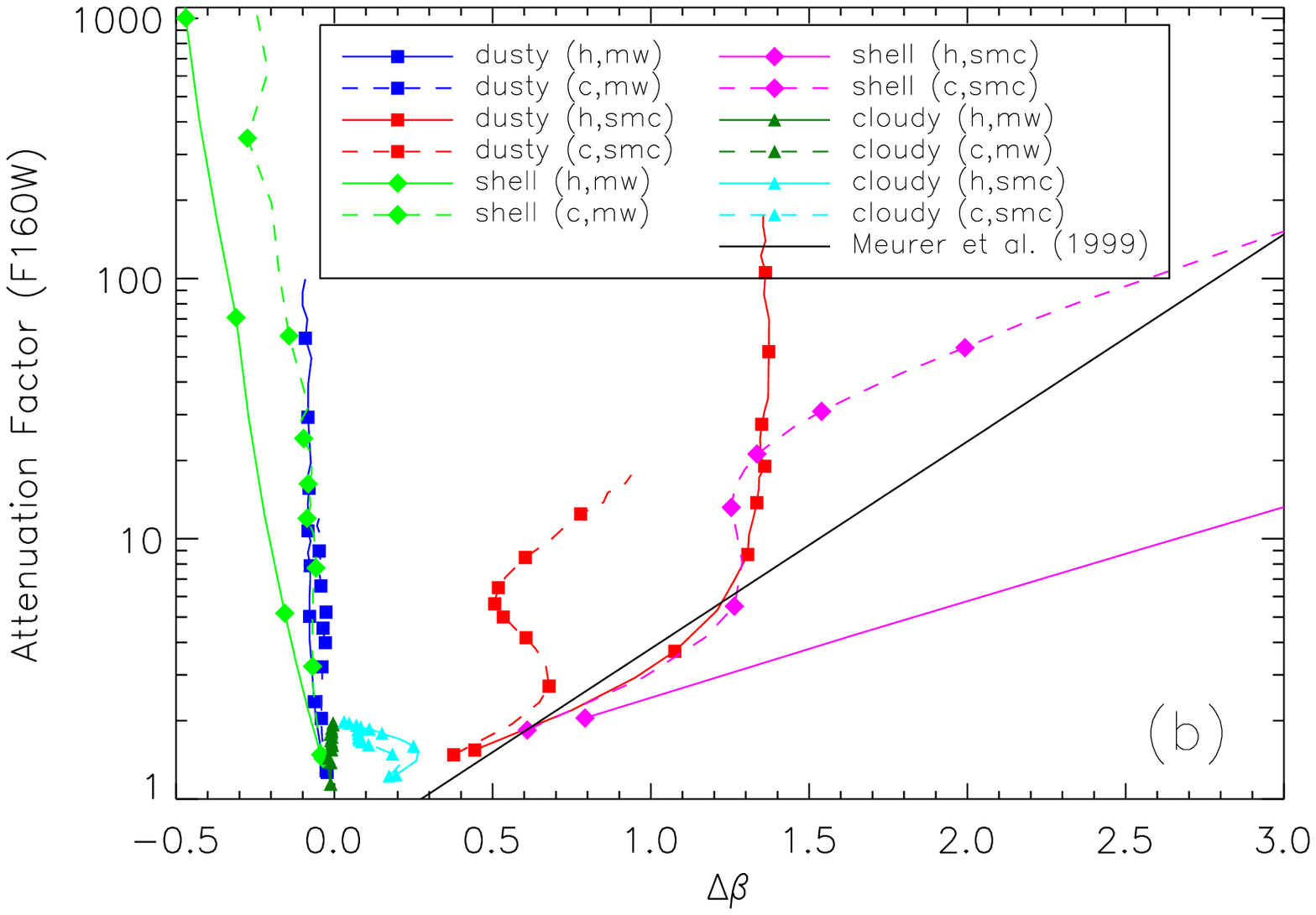}
\caption{The attenuation factor in the HST/WFPC2 F160BW filter is
plotted versus $E(B-V)$ (a) and $\Delta\beta$ (b). The symbols are
plotted every third model $\tau_V$ point. \label{fig_A1600}}
\end{figure*}

Fortunately, our models provide an excellent set of tools with which
to evaluate different approaches to the problem of determining the
far-UV attenuation factor. In Fig.~\ref{fig_A1600}a we illustrate the
computed attenuation factor at 1600~\AA\ as a function of $E(B-V)$
found for different model geometries and dust types. As anticipated
above, the observation that most observed values of $E(B-V)$ for
Lyman-break galaxies fall into the narrow range of 0.2 to 0.3 barely
constrains the attenuation factor to the range 4 to 200.  This is not
very useful as a constraint.  In Fig.~\ref{fig_A1600}b we investigate
the use of the spectral index $\beta$ as a tool to evaluate the
attenuation factor. Our radiative transfer models do not assume a
specific energy distribution for the source spectrum. Hence, we
predict only the change in the spectral index $\beta$ which occurs as
a result of the wavelength dependent attenuation. The GALEX program
advocates this particular approach by suggesting that the change in
the UV slope of the galaxies' SED can be converted into information
about the applicable attenuation factor. Meurer et al.\ (1999) have
taken the additional step of calibrating the correlation between the
spectral index $\beta$ and the UV-attenuation, derived for nearby
UV-selected starburst galaxies.  In observed starburst spectra, this
change in spectral index ranges from zero to 3 (Meurer et al.\
1997). Our Fig.~\ref{fig_A1600}b reveals that correlations between the
spectral index change and the UV attenuation factor are extremely
model dependent and even more dependent on the type of dust present in
a given system. The calibration proposed by Meurer et al.\ (1999) is
shown in Fig.~\ref{fig_A1600}b as a solid black line. For this to be
valid universally, one must assume that SMC-type dust is present in
all starburst galaxies and that SHELL-type geometries with various
degrees of clumpiness are the prevailing star/dust geometries.  Even
when restricting considerations to SMC-type dust, the range of
observed values of $\Delta\beta$ is consistent with attenuation
factors in the range 0--150. As is evident from Fig.~\ref{fig_A1600}b,
MW-type dust is not expected to produce positive values of
$\Delta\beta$. This is contrary to observations in starbursts, where
such positive changes in the UV spectral index are seen (Meurer et
al.\ 1999). However, MW-type dust will reveal its presence by the
appearance of a 2175 \AA\ feature. This has been seen in the heavily
obscured starburst nucleus of M33 (\cite{gor99a}), which indicates
that SMC-type dust is not a universal characteristic of all starburst
environments.  However, the preference for SMC-type dust in starburst
galaxies is affirmed, when we compare values of $F_{\rm FIR}/F_{\rm
F160BW}$ flux ratios and corresponding values of $\Delta\beta$ from
observations (Meurer et al.\ 1997) with our models in
Fig.~\ref{fig_UVratio}. For this calculation we assumed that the
unattenuated source SED was given by a 1 or a 1000 Myr constant star
formation SED of Fioc \& Rocca-Volmerange (1999).  The observed data
points clearly follow the trend set by the model for clumpy SHELL
configurations with SMC dust. This agreement would become even more
convincing, were we to present models with different ratios of clump
to interclump density, different clump size, and different filling
factors, which is, however, beyond the scope of this present
discussion. The demonstrated model dependence of the ratio of FIR to
F160BW flux suggests that the observed scatter in the data points is
most likely real, as was concluded by Meurer et al.\ (1999) also. The
spread in the FIR to F160BW ratio for a given value of $\Delta\beta$
corresponds to a factor of 10 uncertainty in the corresponding UV
attenuation factor. We conclude, therfore, that the UV spectral slope
is not a reliable indicator of the dust attenuation in a given
galaxy. It is usable only, if a large sample of starburst galaxies
with relatively uniform characteristics is being considered at once.

\begin{figure*}[tbp]
\plottwo{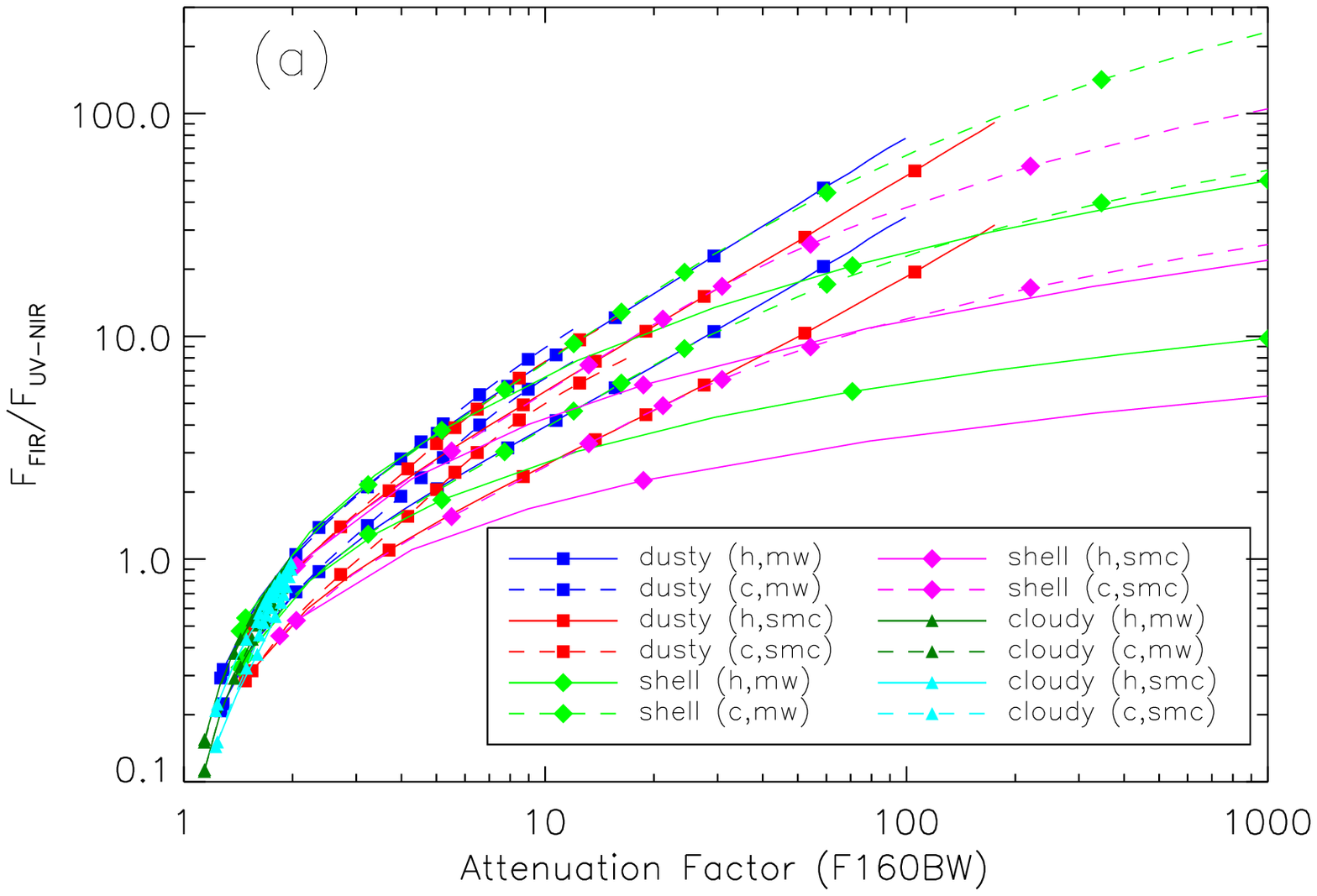}{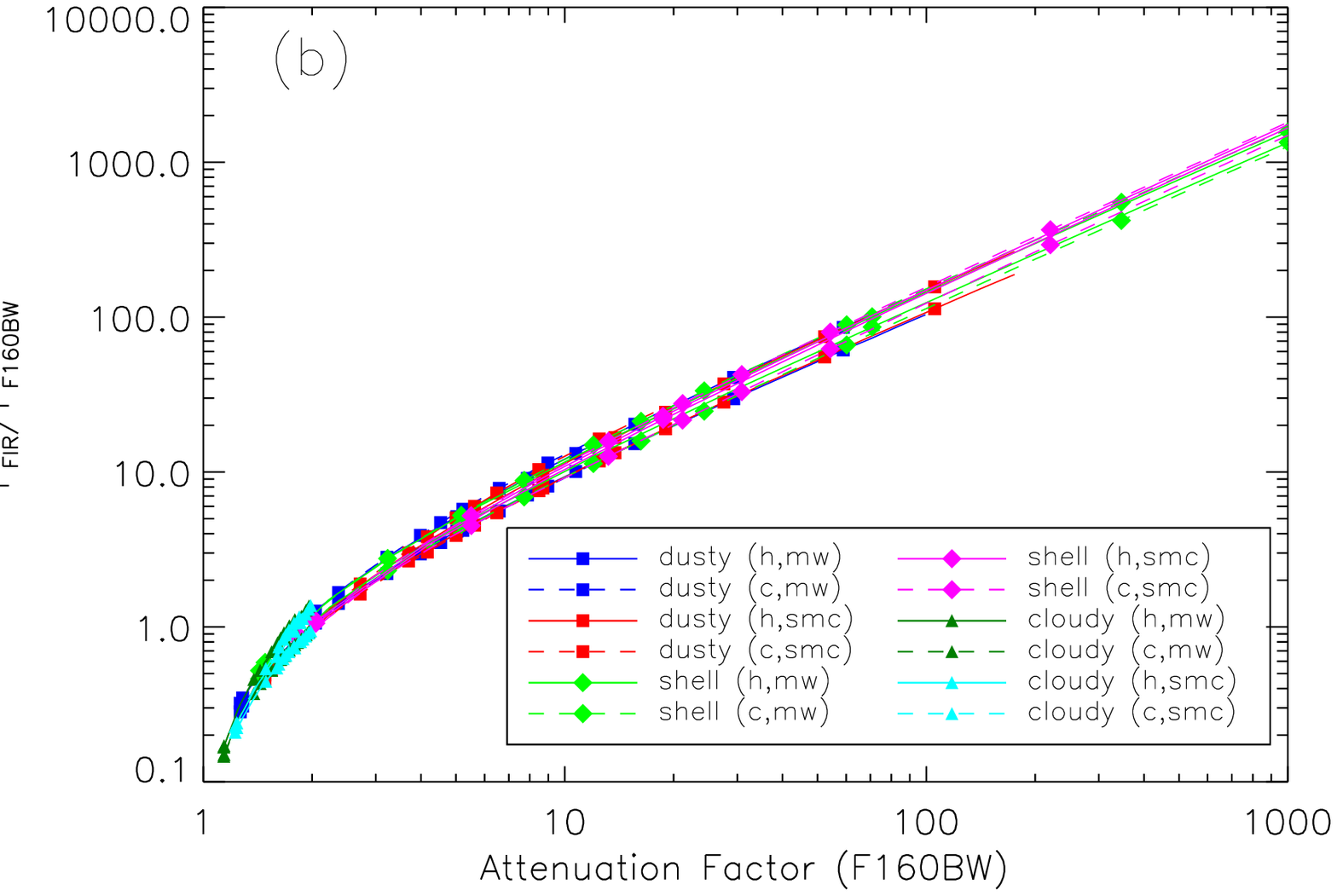}
\caption{The far-IR to UV-NIR (a) and far-IR to F160W (b) flux ratios
are plotted versus the attenuation factor at 1600 \AA.  Each model
curve is shown twice, once assuming a 1 million year old constant star
formation SED and once assuming a 1 billion year old constant star
formation SED.  The symbols are plotted every third model $\tau_V$
point. \label{fig_ratio_1600} }
\end{figure*}

In Fig.~\ref{fig_ratio_1600}a, we plot the ratio of dust-emitted IR
flux to the integrated UV--NIR flux against the attenuation factor at
1600~\AA. The model calculations assume two starbursts of different
ages, one a 1 million year old constant star formation SED, the other
a 1 billion year old constant star formation SED.  We note that even
this ratio does not yield a unique relation to the UV attenuation
factor. In particular, clumpiness in a given dust distribution will
reduce the attenuation at 1600 \AA\ for a given value of the ratio of
dust-emitted IR flux to the integrated UV--NIR flux, because
clumpiness is associated with a much grayer attenuation function. This
is particularly evident in SHELL-type geometries. Variations in the
age of the starburst and the associated changes in the UV--NIR SED
affect the ratio as well, leading to a significant model dependence of
the predicted attenuation factor at 1600~\AA. By contrast, if we plot
the ratio of integrated FIR flux to the measured flux at 1600~\AA\
against the attenuation factor at 1600~\AA, as shown in
Fig.~\ref{fig_ratio_1600}b, the model and age dependence
practically vanish. We conclude, therefore, that the UV attenuation
factor can be evaluated reliably for individual galaxies only, if a
complete spectral survey of the FIR emission of the galaxy is carried
out and a flux measurement in the far-UV is obtained.
 
\vspace*{0.05in}
\epsscale{0.45}
\begin{center}
\plotone{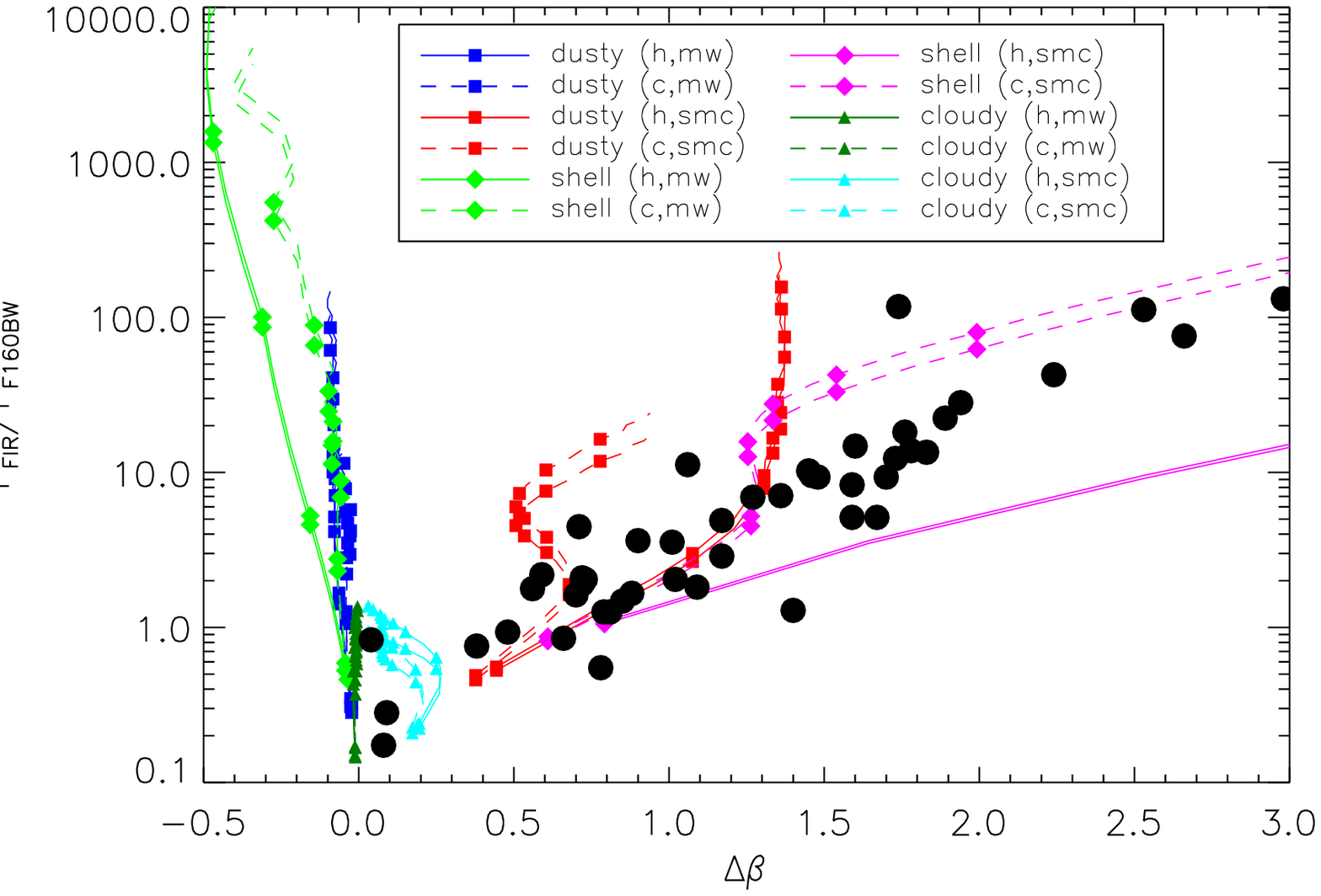}
\end{center}
\figcaption{The ratio of FIR to F160BW flux ($F_{\rm FIR}/F_{\rm
F160BW}$) is plotted versus $\Delta\beta$.  Each model curve is shown
twice, once assuming a 1 million year old constant star formation SED
and once assuming a 1 billion year old constant star formation SED.
The symbols are plotted every third model $\tau_V$ point.  The solid
black points are from Meurer et al.\ (1999) and $F_{\rm F160BW} \sim
F_{\rm 1600} = \lambda_c f_{\rm 1600}$ and $\lambda_c = 1600$~\AA.
The $\Delta\beta$ values for the galaxies were computed using
$\Delta\beta = \beta + 2.5$. \label{fig_UVratio}}

\subsection{Lack of 2175 \AA\ Extinction Bump in Starbursts \label{sec_bump}} 

Gordon et al.\ (1997) showed that the absence of the 2175~\AA\
extinction bump in the attenuation function of starbursts cannot be
explained by radiative transfer effects but requires the actual
absence of the absorber giving rise to this feature.  This is despite
the fact that some weakening of the 2175~\AA\ feature was seen in the
attenuation curves (Fig.~\ref{fig_ext_mw}) for embedded geometries
involving MW dust and further weakening occurred when clumpiness was
included.  As was found by Gordon et al.\ (1997), the 2175~\AA\ bump
can be suppressed using radiative transfer effects, but at the expense
of a gray extinction throughout the UV.  Thus, the existence of a
substantial slope in the Calzetti (1997) attenuation curve
(Fig.~\ref{fig_ext_cal}) implies that the dust in starburst galaxies
must lack the absorbers which produce the 2175~\AA\ bump.  The dust in
starbursts most closely resembles that found in the SMC
(\cite{gor98}). Recent modeling of this dust by Zubko (1999) shows
that such dust lacks the component normally introduced to produce the
2175~\AA\ feature (graphite), and that the overall size distribution
of SMC grains is shifted toward smaller sizes compared to MW dust.

We find the color index ($F170W - F218W$) to be particularly sensitive
for distinguishing between the presence of SMC dust and MW dust, one
without the 2175~\AA\ feature and the other with. In
Fig.~\ref{fig_f170wf218w_bv}, we show a color-color diagram of
starburst data, which are compared to the colors of unreddened
starburst models (open symbols) and to reddening lines produced by our
models. These reddening lines are attached to different unreddened
models to show them distinctly; they may be shifted horizontally to
fit the data.  Several facts emerge clearly: All starburst data appear
on or below the line of unreddened burst models, i.e. they are
reddened in the ($F170W - F218W$) index.  Only the reddening lines for
SMC dust fall below the unreddened bursts, while all MW models fall
above the unreddened bursts, where no data points are found. Both the
amount of reddening in the ($F170W - F218W$) index and the
distribution of data points in the color plane can best be fitted with
a clumpy SHELL model with SMC dust. We take this (as did Gordon et
al.\ 1997) to be one of the strongest indications that nearby
starbursts contain SMC type dust and that the prevailing geometry is
that of a clumpy SHELL model. This sensitive test might be applied to
the rest-frame UV SEDs of high-redshift galaxies in general in order
to test for the presence of the type dust (MW or SMC).

\vspace*{0.05in}
\epsscale{0.45}
\begin{center}
\plotone{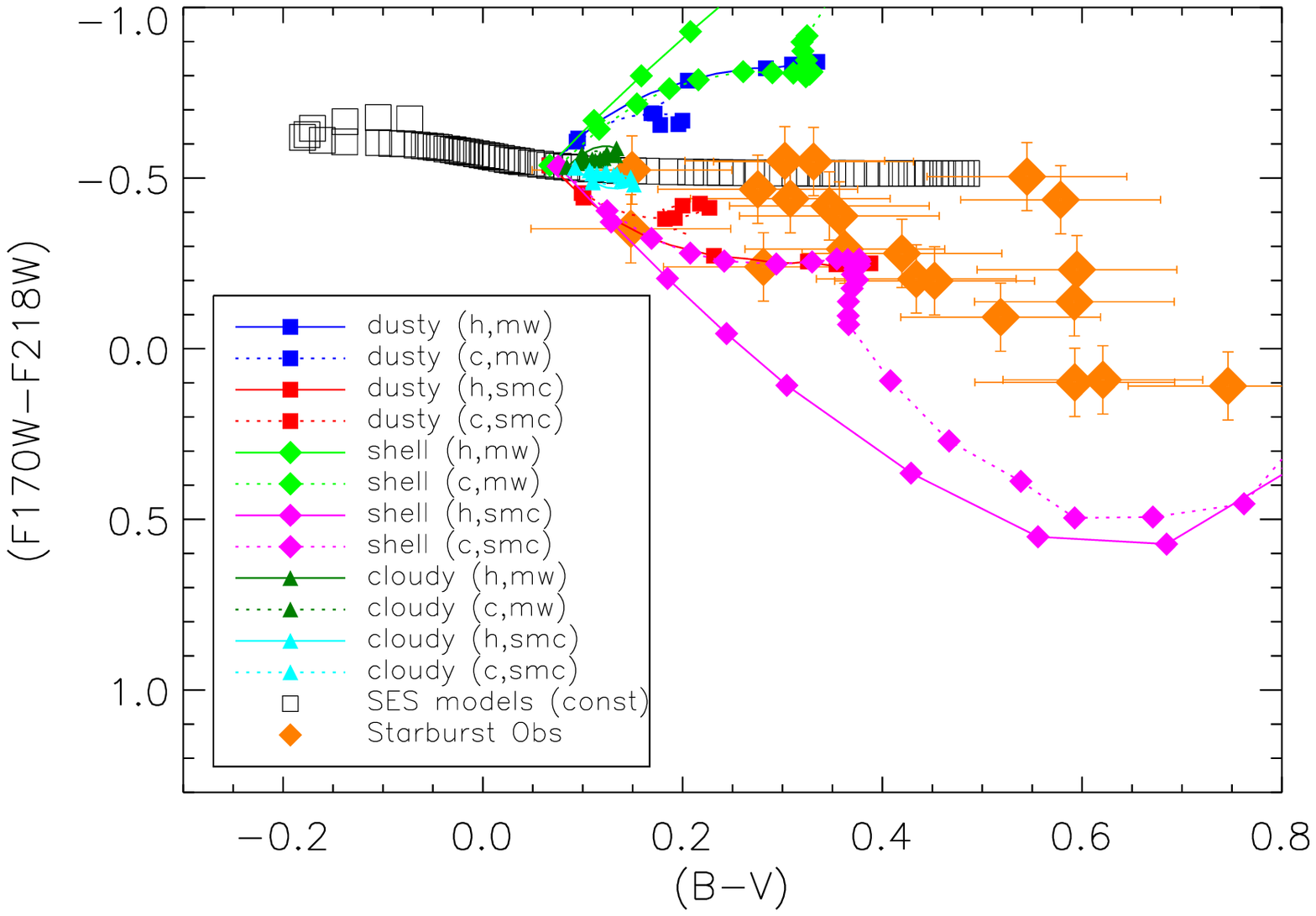}
\end{center}
\figcaption{The (F170W-F218W) and (B-V) colors for starburst observations
and constant star formation SES models are plotted versus each other
along with the reddening trajectories predicted from our radiative
transfer model.  Each reddening trajectory could be attached to any of
the SES model points.  For this figure, we have arbitrarily attached
placed the reddening trajectories to define the envelope of the
starburst observations.  The starburst observations all lie above
[(F170W-F218W) $> -0.6$] the SES models, which is only possible if the
dust in starburst galaxies is similar to that found in the
SMC. \label{fig_f170wf218w_bv}}
\vspace*{0.05in}

\section{Summary and Conclusions} 

We have presented new multiple-scattering radiative transfer
calculations for three representative types of galactic environments
and two dust types, as found within the Local Group in the form of the
average MW and SMC type dust. Both homogeneous and two-phase clumpy
dust distributions have been explored over a large range of dust
column densities, measured by $\tau_{\rm V} = 0 - 50$. The full
numerical details of the models are made available to the community
via the authors, the authors WWW site, and eventually various
national/international data centers.  The environments examined
include DUSTY, a fully mixed spherical distribution of dust and
embedded stars; SHELL, a dust-free star filled region, surrounded by a
dust shell; and CLOUDY, a mixed dust-star region surrounded by stars
in a dust-free region on the outside.  The models cover the wavelength
range 1000
\AA\ to 30000 \AA\ with exceptionally close coverage of the
rest-frame UV. This aspect makes these models particularly applicable
to UV-selected starburst galaxies and Lyman-break galaxies.

We examined the models with respect to predicted reddening effects,
attenuation functions, and the contribution of scattered light to the
total flux as a function of wavelength. All model types except the
homogeneous SHELL model exhibit reddening saturation effects. In
particular, models that are likely representatives for starburst
galaxies show values of $E(B-V)$ not in excess of 0.3, despite a large
range of dust column density. Reddening ``vectors'' in color-color
diagrams are complex non-linear functions depending on the prevailing
dust type and structure, as well as on the geometry.  Attenuation
functions, which measure the wavelength-dependent reduction in the
total stellar and scattered flux from a dusty galaxy, exhibit a trend
toward increasing grayness with increasing dust column density. This
trend is stronger in clumpy dust distributions than in homogeneous
ones. No justification was found for the use of a universal
attenuation function for the analysis of a large sample of galaxies.
In particular, we found that the widely-employed ``Calzetti Law'' is
most closely reproduced by our clumpy SHELL model with SMC-type dust
and a dust column density corresponding to $\tau_V = 1.5$. Such a case
corresponds to an attenuation optical depth in the UV of $\tau_{\rm
att} = 2.0$ or a UV attenuation correction factor of 7.4. The Calzetti
law is valid for statistical samples which have similar dust/star
geometries as found in UV-selected starbursts and which have dust
column densities within narrow limits. Under no circumstances should
it be used for the attenuation correction for single galaxies.

When investigating various approaches to determining this attenuation
factor in general, we found that the use of the color excess $E(B-V)$
and the UV spectral index $\beta$ lead to results that depend strongly
on dust type, dust distribution, and source-dust geometry. Only the
measurement of the $F_{\rm FIR}/F_{\rm F160BW}$ flux ratio promises
reasonable certainty for the determination of the UV attenuation
correction factor in individual galaxies. This reaffirms the need for
observations of starburst galaxies and Lyman-break galaxies over the
entire range of rest-wavelengths, particularly the FIR, in addition to
the UV.  We further showed that observations can clearly discriminate
between the presence of MW-type dust and SMC-type dust through the use
of a $(F170W - F218W)$ vs.\ $(B-V)$ color-color diagram. When applied
to existing data on UV-selected nearby starburst galaxies, a
consistent prevalence of SMC dust is found.

We gratefully acknowledge support from NASA grants NAG 5-3367 and
NAGW-3168 to The University of Toledo. Constructive conversations with
Daniela Calzetti, Geoffrey Clayton, and Karl Misselt are also
gratefully acknowledged.

\end{document}